\documentclass[lettersize,journal]{IEEEtran}
\usepackage{amsmath,amsfonts}
\usepackage{algorithmic}
\usepackage{algorithm}
\usepackage{caption}
\usepackage{array}
\usepackage{pdfpages}
\usepackage{xcolor}
\usepackage[caption=false,font=normalsize,labelfont=sf,textfont=sf]{subfig}
\usepackage{textcomp}
\usepackage{stfloats}
\usepackage{url}
\usepackage{verbatim}
\usepackage{graphicx}
\usepackage{svg}
\usepackage{cite}
\usepackage{multirow}
\usepackage{makecell}
\usepackage{tabularx}
\usepackage{booktabs}
\usepackage{hyperref}
\usepackage[table]{xcolor} 
\definecolor{mygreen}{RGB}{209, 233, 203} 
\definecolor{mygray}{RGB}{235, 235, 235}

\newcommand{\rev}[1]{\textcolor{black}{#1}}
\newcommand{\revv}[1]{\textcolor{black}{#1}}
\newcommand{\revvv}[1]{\textcolor{black}{#1}}
\newcommand{\HVmeanstd}[2]{%
  \makecell[c]{%
    #1\\[-1pt]
    $\pm\,$#2%
  }%
}

\newcommand{\HVbestmeanstd}[3]{%
  \makecell[c]{%
    \textbf{#1}$^{#3}$\\[-1pt]
    $\pm\,$\textbf{#2}%
  }%
}
\begin{document}

\title{Efficient Parameter Calibration of Numerical Weather Prediction Models via Evolutionary Sequential Transfer Optimization}

\author{Heping Fang, Bingdong Li, and Peng Yang,~\IEEEmembership{Senior Member,~IEEE}
\thanks{This work was supported in part by the National Natural Science Foundation of China under Grants 62272210 and 62331014, in part by the AI for Science Program of the Shanghai Municipal Commission of Economy and Informatization under Grant 2025-GZL-RGZN-BTBX-01014. (Corresponding author: Peng Yang)}
\thanks{Heping Fang and Peng Yang are with the Department of Statistics and Data Science, Southern University of Science and Technology, Shenzhen 518055, China. Peng Yang is also with the Guangdong Provincial Key Laboratory of Brain-Inspired Intelligent Computation, Department of Computer Science and Engineering, at the same university.(e-mail: 12331106@mail.sustech.edu.cn; yangp@sustech.edu.cn) Bingdong Li is with the Shanghai Institute of AI for Education, East China Normal University, Shanghai 200062, China (e-mail: bdli@cs.ecnu.edu.cn).}}

\maketitle

\begin{abstract}
The configuration of physical parameterization schemes in Numerical Weather Prediction (NWP) models plays a critical role in determining the accuracy of the forecast. However, existing parameter calibration methods typically treat each calibration task as an isolated optimization problem. This approach suffers from prohibitive computational costs and necessitates performing iterative searches from scratch for each task, leading to low efficiency in sequential calibration scenarios. To address this issue, we propose the SEquential Evolutionary Transfer Optimization (SEETO) algorithm driven by the representations of the meteorological state.
First, to accurately measure the physical similarity between calibration tasks, a meteorological state representation extractor is introduced to map high-dimensional meteorological fields into latent representations. 
Second, given the similarity in the latent space, a bi-level adaptive knowledge transfer mechanism is designed. At the solution level, superior populations from similar historical tasks are reused to achieve a "warm start" for optimization. At the model level, an ensemble surrogate model based on source task data is constructed to assist the search, employing an adaptive weighting mechanism to dynamically balance the contributions of source domain knowledge and target domain data.
Experiments on multiple calibration tasks with varying source--target similarities demonstrate that SEETO consistently improves early-stage calibration efficiency under limited expensive evaluation budgets, while maintaining competitive overall optimization performance.
 \revv{This provides a practical approach for efficient automated calibration of NWP model parameters.}
\end{abstract}


\section{Introduction}
\IEEEPARstart{A}{ccurate} weather forecasting is essential for many real-world applications, including transportation, energy management, and disaster mitigation. Current forecasting methods mainly include Numerical Weather Prediction (NWP) and deep learning approaches \cite{ref5}. Although deep learning has shown promise in data-driven forecasting, NWP remains indispensable because of its physical consistency and reliability, especially for operational forecasting and extreme weather prediction \cite{ref6}. Therefore, improving the forecast accuracy of NWP models remains a central problem in weather forecasting research \cite{ref8}.

NWP models are primarily composed of a dynamical core and physical parameterization schemes. These models integrate multiple physical parameterization schemes (e.g., microphysics, planetary boundary layer, and cumulus convection)  to meticulously characterize the complex physical processes involved in the evolution of atmospheric states. 
Note that, the extensive array of parameter values embedded within these schemes significantly influences the accuracy of the final forecast. In practice, researchers typically adopt the default parameter configurations of these schemes. 
Yet, substantial research indicates that these default settings do not yield optimal forecast results across all geographical regions and weather conditions \cite{ref10}. 
Consequently, to enhance forecast precision, error correction of forecast results is required, and calibrating these key parameters for specific weather scenarios has been essential in improving NWP performance \rev{~\cite{ref52}}.

The goal of the calibration process is to align the output of complex models with real-world observations. This is achieved by adjusting the model parameters. This has been widely studied in various fields, such as meteorology and finance\cite{ref12,ref13}. Existing research typically formulates this task as a computationally intensive multi-objective optimization problem. Currently, evolutionary multi-objective optimization algorithms \cite{ref15} are the primary approach to solving these problems. However, this approach faces a critical bottleneck. The evaluation of each candidate parameter vector incurs a high computational cost. Furthermore, a "universal" optimal solution does not exist for the physical parameterization schemes of NWP models. Different calibration tasks are influenced by various factors, such as geographic region and meteorological background. Consequently, researchers must execute independent optimization processes for multiple tasks. These processes are typically performed from scratch and are computationally expensive \cite{ref8,ref19,ref20,ref21}. For instance, the calibration of parameters of the Weather Research and Forecasting (WRF) model \cite{ref13} illustrates this challenge. The WRF is a widely recognized open-source mesoscale NWP model. A single run requires extensive numerical integration of nonlinear atmospheric dynamic equations. This process operates on high-resolution spatiotemporal 3D grids and involves complex physical parameterizations. As a result, it consumes a large amount of computational resources and time. Benchmarks indicate that a single 24-hour simulation at a resolution of 2.5 kilometers consumes approximately 1,150 core hours \cite{ref23}. Consequently, efficient calibration strategies are essential. Without them, the prohibitive time overhead prevents meeting the requirements for timely forecasting.

In recent years, Surrogate-Assisted Evolutionary Algorithms (SAEAs) \cite{ref24} have been proposed to address computationally expensive optimization problems. 
The core philosophy of these methods involves employing computationally inexpensive surrogate models to substitute for the expensive evaluations of the actual physical problems, thereby enabling the efficient search for superior solution vectors using only a limited budget of evaluations. 
Although existing SAEAs can mitigate the optimization burden for individual parameter calibration tasks of NWP models, the solution process for each task remains isolated when these methods are strictly applied to multiple calibration scenarios, as previously noted. 
Specifically, each new task initiates its search from a "zero-knowledge" state. This approach, which assumes mutual independence among problems, overlooks a critical reality: real-world problems rarely exist in isolation, and similar tasks often share underlying information that can be exploited. 
When effectively leveraged, such information can significantly enhance problem-solving efficiency. Recent studies have made methodological advancements in this direction, collectively referred to as Evolutionary Transfer Optimization (ETO) \cite{ref25}. 

Although the application of ETO to multiple calibration tasks of NWP models holds substantial promise for enhancing optimization efficiency, the implementation of this concept encounters three primary challenges \cite{ref28}. 
The first challenge is task selection, which aims to identify a subset of source tasks that share similarities with the target task from a comprehensive archive of source tasks. This necessitates a transferability metric tailored to NWP model calibration tasks, capable of accurately quantifying the degree of inter-task similarity. 
The second challenge is knowledge representation, which involves determining an effective formalism for the information acquired from source calibration tasks, thereby explicitly defining the specific objects to be transferred. 
The final challenge lies in the knowledge transfer mechanism. Given the discrepancies in objective function landscapes across different NWP calibration tasks, effective adaptive measures are required to bridge the distribution gap between source and target domains, ensuring that the transferred knowledge can be efficiently utilized during the search process of the target task. 
To address the aforementioned challenges and facilitate the application of ETO to multiple NWP calibration tasks, we propose the SEquential Evolutionary Transfer Optimization algorithm driven by meteorological state representation, termed SEETO. 
The main contributions of this study are summarized as follows:
\begin{itemize}
  \item{\revv{This study investigates the application of ETO to address the challenge of repetitive and computationally expensive evaluations inherent in sequential calibration tasks of NWP models.}
  By leveraging knowledge from source tasks related to the target task, the proposed method significantly enhances calibration efficiency. }
  \item{We propose a novel approach to guide the parameter calibration of NWP based on the meteorological state itself. 
  A meteorological state representation extraction model based on self-supervised learning is developed. 
  The learned representations are utilized to accurately quantify the similarity between the target task and potential source tasks, thereby enabling the intelligent selection of source tasks for knowledge transfer.}
  \item{We design a bi-level adaptive knowledge transfer mechanism. Transfer at the model level is facilitated through the construction of dynamically weighted ensemble surrogate models, while transfer at the solution level is achieved by injecting elite solutions from similar tasks.}
\end{itemize}

The remainder of this paper is organized as follows. 
Section~\ref{Section-II} provides an overview of the problem background and related work. 
Section~\ref{Section-III} elaborates on the proposed SEETO algorithm. 
Section~\ref{Section-IV} presents the detailed experimental design, results, and analysis. 
Finally, Section~\ref{Section-V} concludes the paper and outlines directions for future research.

\begin{figure}[!h]
\centering
\includegraphics[width=2.2in]{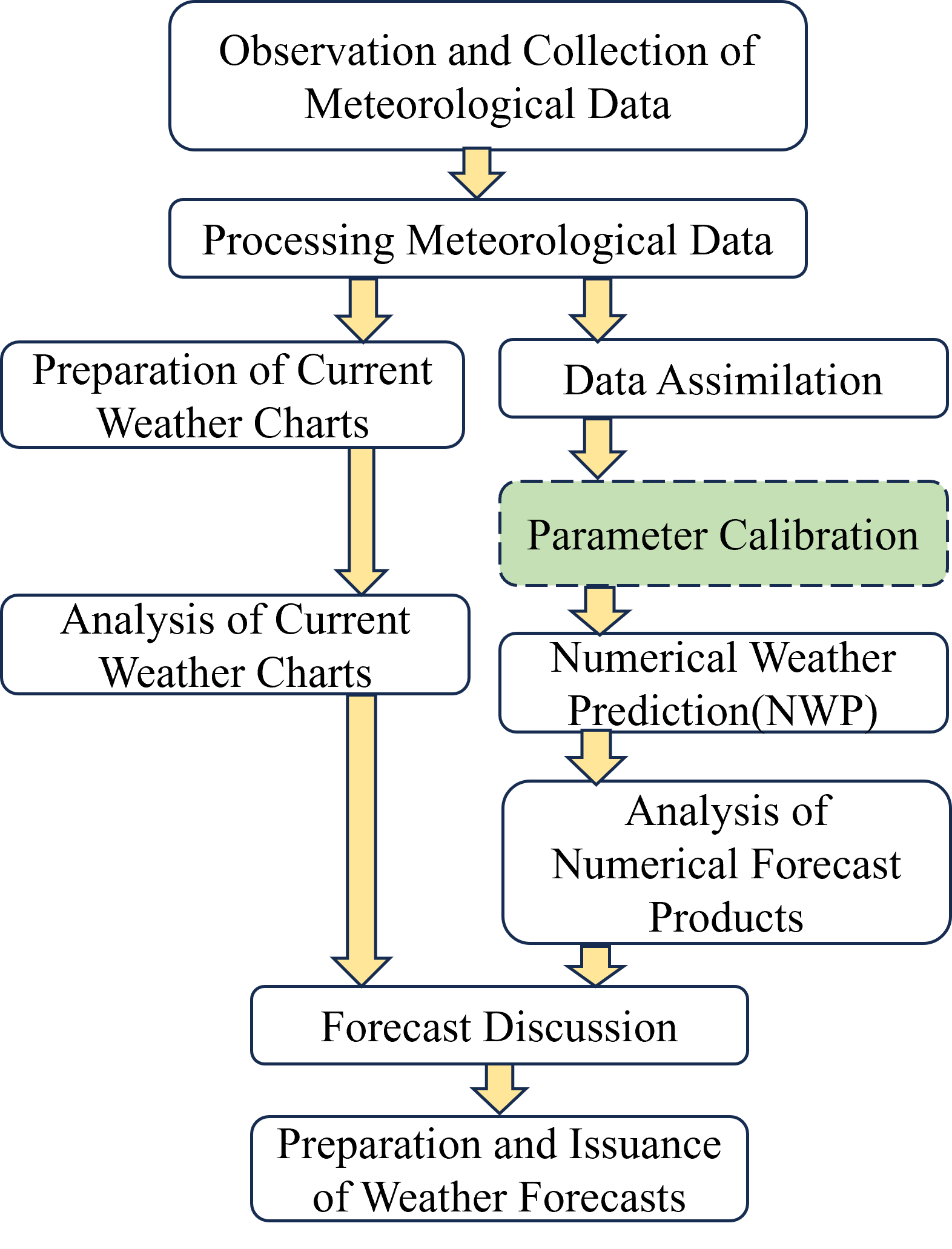}
\caption{Workflow of the Operational weather forecasting.}
\label{fig_1}
\end{figure}

\section{Background And Related Work}
\label{Section-II}
\subsection{Problem Definition}
\revv{Fig.~\ref{fig_1} illustrates the workflow of operational weather forecasting. The process usually begins with the acquisition of meteorological observations from multiple sources, including surface stations, radiosondes, radar, and satellites. The raw data are first quality-controlled, reformatted, and systematically organized, and then split into two branches. The left branch is used to generate diagnostic products such as current weather charts, which describe the actual weather state at the current time and support human forecasters in analyzing the ongoing weather situation. The right branch enters the data assimilation system, where the latest observations are combined with the model background field from the previous forecast to produce an initial analysis field that is as close as possible to the true atmospheric state \rev{~\cite{ref51}}. Based on this analysis field, the NWP model is integrated forward to generate forecasts for the coming period. Therefore, the right branch represents the evolution of weather states in the numerical forecasting process. Parameter calibration is performed before NWP. Its purpose is to optimize key parameters in the physical parameterization schemes according to the discrepancy between historical observations and model outputs, so as to provide more suitable parameter settings for subsequent forecasting tasks. After the numerical forecasts are produced, forecasters further combine the analysis of the current weather charts from the left branch with the analysis of numerical forecast products from the right branch to conduct forecast discussions, and finally prepare and issue weather forecasts. From this workflow, it can be seen that parameter calibration is an important step in the operational chain. It must be completed within a limited operational time window and directly affects the quality of numerical forecasts.}

Consequently, a critical question arises regarding how to efficiently determine the optimal configuration of internal parameters within the physical parameterization schemes of NWP models for specific weather forecasting scenarios. 
We formulate this challenge as the NWP model calibration problem, which is described as follows:

Let $\hat{\boldsymbol{s}} = \{\hat{s}_t\}_{t=1}^T$ denote the observed spatiotemporal variation data of multiple meteorological variables recorded over $T$ time steps. 
Correspondingly, the NWP model, denoted as $\mathcal{M}(\boldsymbol{\theta})$, generates predicted spatiotemporal data $\boldsymbol{s} = \{s_t\}_{t=1}^T$ over the same duration. 
The configuration of parameters $\boldsymbol{\theta}$ exerts a direct influence on the generated forecasts. 
Assuming the existence of an optimal parameter set $\boldsymbol{\theta}^*$ such that $\boldsymbol{\theta}^* = \mathcal{M}^{-1}(\hat{\boldsymbol{s}})$, the objective is to determine the value of $\boldsymbol{\theta}^*$. 
In the literature, this is typically approximated by solving the optimization problem formulated as:
\begin{equation}
  \label{equation_1}
  \boldsymbol{\theta}^* = \mathop{\arg\min}\limits_{\boldsymbol{\theta} \in \Omega} D(\hat{\boldsymbol{s}}, \boldsymbol{s} = \mathcal{M}(\boldsymbol{\theta}))
\end{equation}
where $D$ represents a specific distance metric, typically the Root Mean Squared Error (RMSE) used in parameter calibration. 
In this study, we select wind speed and temperature in $\hat{\boldsymbol{s}}$ and $\boldsymbol{s}$ as calibration objectives. By adjusting the WRF parameter vector, we aim to simultaneously minimize the forecast errors of these two meteorological variables.

\begin{figure}[!h]
\centering
\includegraphics[width=3.2in]{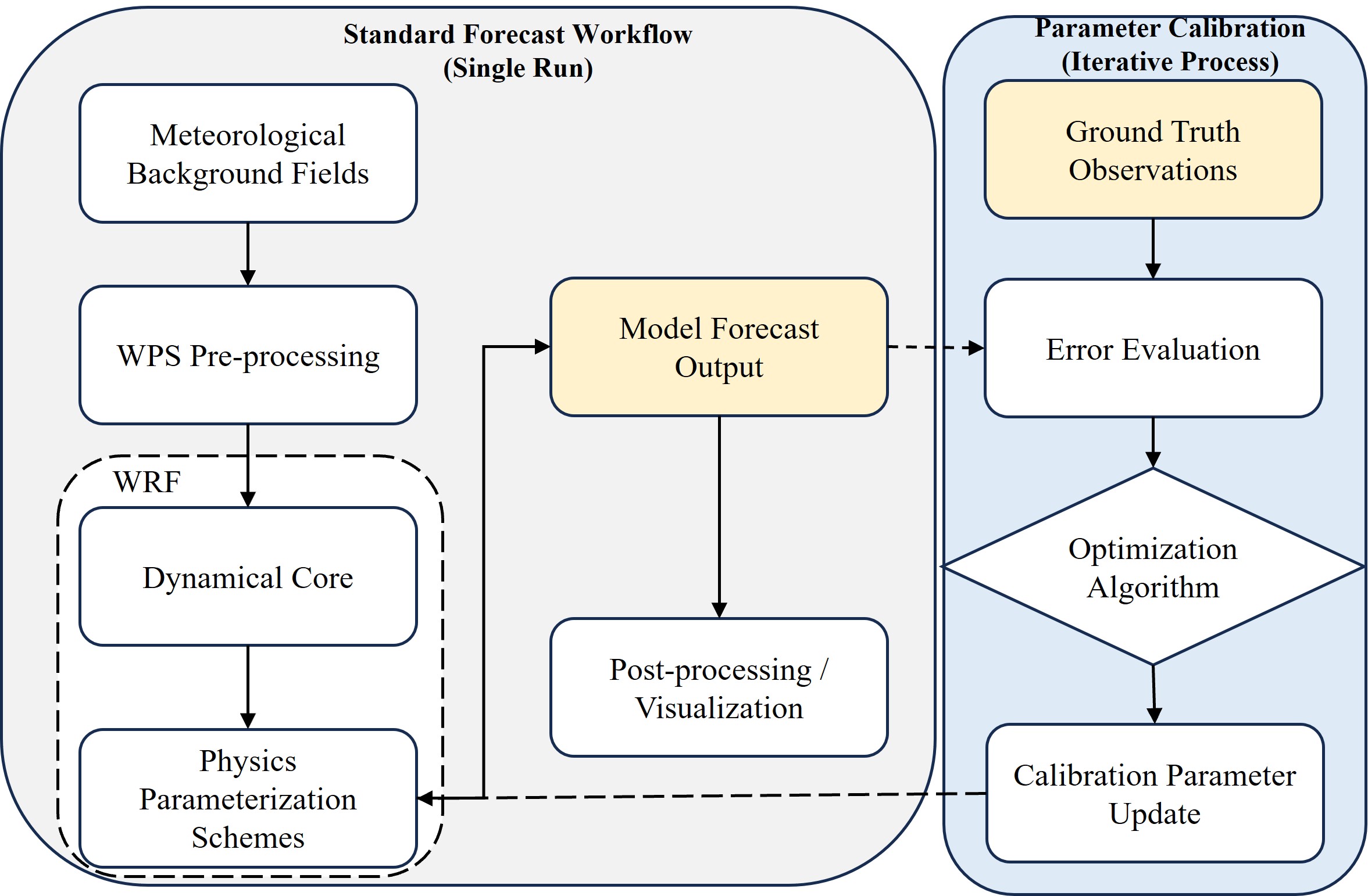}
\caption{Flowchart of the NWP Model Parameter Calibration.}
\label{fig_2}
\end{figure}

\subsection{Calibration of NWP Models}
For parameter calibration, traditional manual tuning based on empirical experience or "trial-and-error" is not only time-consuming and computationally expensive but also struggles to locate the global optimum within a high-dimensional parameter space. To overcome these limitations, the field has comprehensively shifted towards automatic parameter calibration methods \cite{ref8,ref19,ref20,ref21}, The general workflow of such methods is illustrated in Fig.~\ref{fig_2}, where \revv{WPS denotes the WRF Preprocessing System}, and the parameter configurations are dynamically adjusted based on errors between observational data and forecast results.

Evolutionary Algorithms (EAs) have gained significant traction in this domain due to their robust global search capabilities and proficiency in handling multi-objective optimization problems. However, a single simulation of NPW models typically consumes substantial computational resources and time. To mitigate the bottleneck of exorbitant evaluation costs, SAEAs have been introduced. These approaches utilize computationally inexpensive mathematical models, such as Gaussian Processes (GP) or Neural Networks, to approximate expensive NWP simulations, thereby assisting the algorithm in rapidly converging towards the Pareto optimal front with a limited budget of real evaluations.
In recent years, SAEAs have achieved significant success in WRF parameter calibration. For instance, Chinta et al. \cite{ref20} proposed the Multi-Objective Adaptive Surrogate Model-based Optimization (MOASMO) algorithm, which successfully optimized WRF model parameters and significantly reduced prediction errors for four key meteorological variables during high-intensity rainfall events in the core region of the Indian summer monsoon. Similarly, focusing on the meteorological characteristics of the Beijing area, Wang et al. \cite{ref19} developed a Knee Point-based Multi-objective Optimization (KMO) algorithm that effectively reduced the number of WRF evaluations via Gaussian Process surrogates, achieving dual improvements in precipitation and temperature forecast accuracy.

Despite the fact that the aforementioned surrogate-based methods have alleviated computational pressures to a certain extent, they suffer from a fundamental limitation: the lack of a mechanism for reusing historical knowledge. When maximizing performance for each new task, these algorithms are compelled to initialize populations randomly and train surrogate models from scratch. This "amnesic" search paradigm neglects the physical continuity and similarity of meteorological systems across spatiotemporal distributions, forcing the algorithm to expend valuable computational resources on blind exploration during the initial optimization phase, thus making it difficult to achieve an efficient "warm start."

\subsection{Evolutionary Transfer Optimization}
\revv{Recent advancements in meta-learning, and ETO indicate that the paradigm of problem-solving is shifting towards knowledge reuse, underscoring the potential of leveraging commonalities among similar tasks. The core idea of ETO is to transfer useful knowledge among multiple related optimization tasks, thereby improving the search efficiency and solution quality of the target task. For the NWP parameter calibration problem investigated in this study, different calibration tasks are not mutually independent: as operational rolling forecasts and historical calibration processes continue to advance, existing tasks are progressively accumulated to form a reusable historical knowledge repository, while newly arriving target tasks must be calibrated under a strictly limited budget of expensive evaluations. Therefore, this setting constitutes a sequentially arriving, multi-task, computationally expensive multi-objective NWP parameter calibration scenario, which is more consistent with a sequential transfer optimization problem oriented towards practical operational rolling calibration requirements.}
The mathematical formulation of such problems is described as follows:
\begin{equation}
  \label{equation_2}
  \min_{x \in \Omega} F(x) = \min_{x \in \Omega} \left( f_1(x \mid \mathcal{H}), \dots, f_m(x \mid \mathcal{H}) \right)
\end{equation} 
Here, $F$ denotes the target calibration task, where $m$ determines the number of calibration objectives. $x$ represents the decision vector within the decision space $\Omega$, and $\mathcal{H}$ serves as a knowledge base that contains available information from source tasks. Specifically, it is assumed that $\mathcal{H}$ encapsulates the search experience derived from $K$ source tasks, $F_1, F_2, \ldots, F_K$. The ETO method uses valid knowledge from $\mathcal{H}$ to accelerate the search process for the target task $F$.

\revv{From the perspective of research directions, the task scenario studied in this work is closely related to three research lines.} \revv{\textbf{Multi-source transfer optimization \cite{ref29,ref30}:} Existing studies focus on selecting useful source tasks and transferring knowledge. Among them, induction-driven methods \cite{ref31} adjust source task weights using post-transfer online feedback, whereas analogy-driven methods \cite{ref32} support "warm start" by precomputing task similarity. Our method belongs to the latter category. However, unlike general approaches that rely on early target task sampling or decision-space statistical distances, we directly measure the physical similarity between source and target tasks in the latent space of meteorological state representations, thereby enabling source task retrieval and weight assignment without target side sampling.}

\begin{figure*}[!t]
\centering
\includegraphics[width=6.2in]{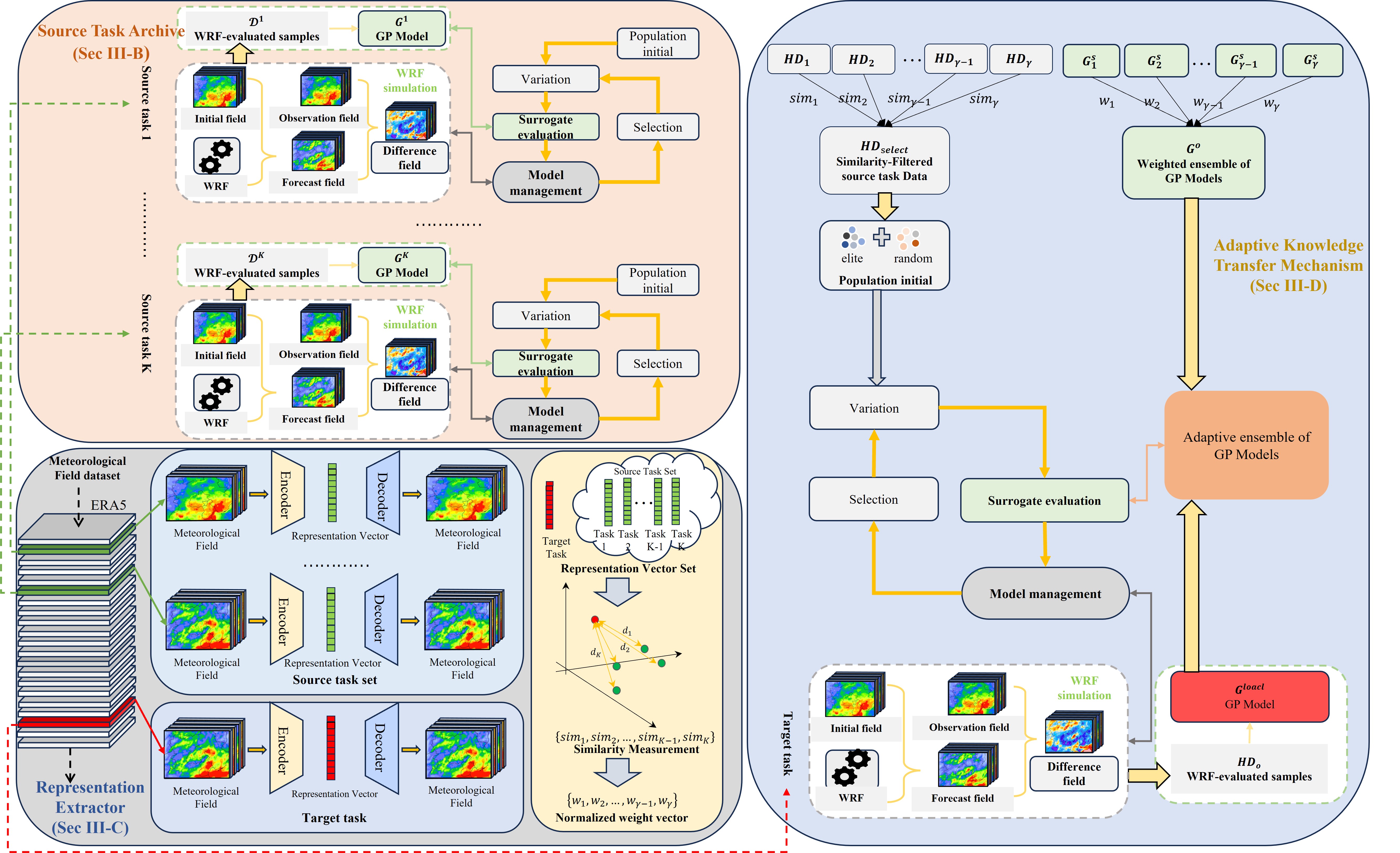}
\caption{The framework of the SEETO.}
\label{fig_3}
\end{figure*}

\revv{\textbf{Evolutionary sequential transfer optimization (ESTO) \cite{ref28,ref33}:} Previous studies mainly investigate how historical experience can be reused to accelerate the optimization of newly arriving tasks when tasks arrive sequentially over time. Common forms include algorithm-based \cite{ref34}, solution-based \cite{ref33,ref36}, and model-based methods \cite{ref37,ref38}. In this work, we instantiate this idea in the multi-task NWP parameter calibration scenario and design a bi-level adaptive knowledge transfer mechanism for extremely limited expensive evaluation budgets: elite solution injection at the solution level for warm start, and adaptive fusion of multi-source surrogates with the target task local surrogate at the model level.}

\revv{\textbf{Multi-problem surrogate modeling \cite{ref38}:} Existing methods usually utilize data or surrogate models from multiple related problems to assist the modeling of the current problem, thereby reducing the number of real evaluations. Unlike these approaches, our method does not depend on early target task sampling to estimate cross-task relationships. Instead, it uses meteorological state representations to drive multi-source surrogate ensemble and adaptive fusion with the target task local surrogate. In addition, we further analyze the empirical behavior of the method under high- and low-similarity task regimes, in order to reveal how transfer gains and negative transfer risks vary with task similarity.}

\revv{From the perspective of ETO components, existing methods typically include an evolutionary search core, a similarity measurement module, a transfer strategy module, a state representation or feature extraction module, and a knowledge base module. SEETO follows this general framework, but its key differences can be understood from the origin of its components, namely, \textbf{a newly introduced component} and \textbf{two adaptively refined components built upon existing ETO modules}. First, we introduce a meteorological state representation extractor as a new component to derive physically meaningful task representations from high-dimensional meteorological fields, rather than relying directly on statistical discrepancies in the decision or objective space. Second, we adapt the similarity measurement module by building it on top of these meteorological representations, so that it can retrieve the most relevant source tasks from the knowledge base and support subsequent elite solution transfer and surrogate ensemble. Finally, we adapt the transfer strategy module by designing a bi-level adaptive knowledge transfer mechanism tailored to extremely limited expensive evaluations: at the solution level, weighted elite solution injection from multiple similar source tasks provides a warm start. At the model level, dynamic fusion of multi-source surrogates and the target-task local surrogate guides the search process.}

\revvv{Beyond ETO, algorithm configuration and hyperparameter optimization address the broader problem of expensive search under limited evaluation budgets. F-Race \cite{birattari2010f} and irace \cite{lopez2016irace} use statistical racing to progressively eliminate statistically inferior configurations. In contrast, Hyperband \cite{li2018hyperband} use multi-fidelity scheduling to allocate more resources to promising configurations. In Bayesian optimization, MOTPE \cite{ozaki2020multiobjective} supports sample-efficient multiobjective search, whereas RGPE \cite{feurer2018scalable} and two-stage transfer surrogate models enable cross-task transfer.Although the aforementioned methods are not the primary focus of this study, they provide complementary perspectives and help position our work more clearly within the broader literature.}

\section{Proposed Algorithm}
\label{Section-III}
\subsection{Framework}
\revv{Fig.~\ref{fig_3} illustrates the overall architecture of SEETO, while Algorithm~\ref{alg:seeto} formalizes the execution flow for each newly arriving target task.}

\revv{First, to alleviate the "cold start" problem during the initial stage of optimization, the algorithm retrieves similar source tasks from the archive. Specifically, in the first stage (Lines~1--9), the meteorological state representation extractor (Section~III-C) maps high-dimensional meteorological fields into a latent space, where physically consistent source--target similarity measurement and transfer weight assignment are conducted.}
\revv{This retrieved information is subsequently utilized by the bi-level adaptive knowledge transfer mechanism (Section~III-D). In the second stage (Lines~10--18), solution level transfer injects elite solutions from similar source tasks into the target task's initial population to provide a high quality warm start. In the third stage (Lines~19--32), model level transfer constructs a weighted ensemble of source task surrogates and adaptively fuses it with the local surrogate as evaluations accumulate, guiding the surrogate-assisted search under a limited budget of expensive evaluations.}
\revv{Finally, in the fourth stage (Lines~33--35), the meteorological state, evaluated samples, and surrogate model are stored in the archive. This feedback loop enables the knowledge base to progressively accumulate experience over the task sequence, providing increasingly effective transfer support for future tasks.}

\subsection{Source Task Archive}
In scenarios involving multi-task calibration, traditional methodologies typically discard data generated during the optimization of source tasks, thereby necessitating an independent optimization process from scratch for each incoming target task. Consequently, constructing a source task data archive $\revv{{\mathcal{A}}}$ is imperative. As historical calibration tasks accumulate, the repository of transferable knowledge expands, facilitating the rapid adaptation of the optimization algorithm to newly arriving target tasks. The archived data for each source task primarily consists of the following key components:
\begin{itemize}
  \item{{\bf{Data Sources for Representation Extraction}}($\hat{\boldsymbol{s}}^i$): As formulated in Eq. \eqref{equation_1} , excluding the parameter vector $\boldsymbol{\theta}$ to be calibrated, the similarity between a source task and a target task depends on the similarity of their corresponding meteorological states, provided that other NWP model $\mathcal{M}$ configurations remain consistent. 
  Here, a representation extractor is employed to derive a latent representation of the meteorological state for each source task, which serves as the basis for the subsequent calculation of the similarity metric between source and target tasks.}
  \item{{\bf{Set of Evaluated Solutions}}($\mathcal{D}^i$): During the iterative optimization of each source task, the selected parameter vectors $\boldsymbol{\theta}$ are subjected to computationally expensive evaluations via the NWP model $\mathcal{M}$, and the resulting data are progressively collected. 
  Upon completion of the optimization, the accumulated high-fidelity data pairs ($\boldsymbol{\theta}$, $\mathbf{F}$) are screened (details provided in Section III-D) to identify the optimal solution set. 
  Injecting this set into a similar target task accelerates the optimization process.}
  \item{{\bf{Surrogate Model}}($G^i$):Throughout the optimization of each source task, a GP surrogate model is constructed using the evaluated parameter vectors ($\boldsymbol{\theta}$) and their corresponding forecast performance values ($\mathbf{F}$). 
  This model approximates the mapping between parameter settings and model performance. If a target task exhibits high-similarity to a specific source task, the source's surrogate model can be directly leveraged as an auxiliary model. 
  This allows for greater exploration of the optimal solution space under the constraint of limited expensive evaluations.}
\end{itemize}

As the volume of source task data grows, incoming target tasks can evaluate their similarity against a broader array of source tasks, thereby increasing the availability of suitable candidates for effective knowledge transfer.

\subsection{Representation Extractor}
In multi-calibration scenarios, effectively leveraging knowledge from existing source calibration tasks to assist a new target task necessitates the identification of physically similar source tasks. 
Consequently, accurately quantifying the similarity between two tasks emerges as a critical challenge. In the context of parameter calibration, the objective is to minimize the discrepancy between the model output and the known ground truth observed meteorological states. 
Given that the parameter vector $\boldsymbol{\theta}$ serves as the optimization variable while other configurations of the NWP model $\mathcal{M}$ remain consistent, the distinction between tasks is primarily defined by their underlying weather conditions. 
Therefore, it is sufficient to quantify task similarity by measuring the similarity of the observed meteorological state data characterizing each task.

Let the time series of ground truth observed meteorological state data be denoted as $\{\hat{s}_t\}_{t=1}^T$, where $\hat{s}_t \in \mathbb{R}^{\lambda \times H \times W}$. Specifically, ${\hat{s}}^i$ and ${\hat{s}}^o$ denote the meteorological state data for the $i$-th source task and the target task, respectively.
Here, the meteorological state data consists of $\lambda$ selected meteorological elements, with each element serving as a distinct channel. The variables $W$ and $H$ denote the width and height of these meteorological elements, respectively.
Quantifying the similarity of such meteorological state data constitutes a similarity measurement problem for high-dimensional spatiotemporal data. 
Traditional metrics typically compute pixel-level errors point-wise. However, they are susceptible to the "Double Penalty" problem \cite{ref40} and often fail to capture high-dimensional nonlinear spatial features. 

To address this, we employ a deep learning model for representation extraction, mapping high-dimensional meteorological fields into a low-dimensional latent space. 
As illustrated in Eq. \eqref{equation_3}, the task similarity is determined by calculating the average cosine similarity between the corresponding latent representation vectors of the source and target tasks over the time sequence:
\begin{equation}
  \label{equation_3}
  \begin{aligned}
  z_t^i &= f(\omega, \hat{s}_t^i) \\
  z_t^o &= f(\omega, \hat{s}_t^o) \\
  sim^i &= \frac{1}{T} \sum_{t=1}^{T} \frac{z_t^i \cdot z_t^o}{\|z_t^i\| \|z_t^o\|}
  \end{aligned}
\end{equation}
where $\hat{s}_t^i$ and $\hat{s}_t^{o}$ denote the observed meteorological states for the $i$-th source task and the target task at time step $t$, 
respectively. $\omega$ denotes the learnable parameters of the feature extraction network $f(\cdot)$, and $z_t^i$ and $z_t^{o}$ represent the extracted low-dimensional latent representation vectors.

The representation extraction model comprises two components: an encoder and a decoder. As illustrated in Eq.\eqref{equation_4}, the encoder consists of multiple downsampling layers and ResNet layers, 
designed to extract continuous encoded features $z_t$ from the input meteorological state data $\hat{s}_t$, where the downsampling layers are implemented via 2D convolutions. 
The architecture of the decoder is the inverse of the encoder, composed of multiple upsampling layers and ResNet layers. 
Starting from the low-dimensional latent representation vector $z_t$, it progressively reconstructs the original meteorological state data $\hat{s}'_t$, with the upsampling layers realized using 2D transposed convolutions. 
We define the loss function as the reconstruction error between the reconstructed meteorological state $\hat{s}'_t$ and the original input $\hat{s}_t$. 
Given that meteorological state data typically consists of continuous physical quantities, 

\begin{algorithm}[H]
\caption{SEETO}
\label{alg:seeto}
\begin{algorithmic}[1]
\color{black}
\REQUIRE ~\\
     Target meteorological state sequence $\hat{\mathbf{s}}^{o}$; Source archive $\mathcal{A}=\{(\hat{\mathbf{s}}^i,\mathcal{D}^i,G_i)\}_{i=1}^{K}$; Autoencoder-based feature extractor $f$; Population size $n_p$; WRF evaluation budget $FE_{max}$; Transfer parameters $\gamma, \Gamma, \rho, c$.
\ENSURE 
    $PS^*$

{\fontsize{9}{12.5}\selectfont\texttt{// Stage 1: Retrieve similar source tasks by meteorological state similarity}}
\STATE \parbox[t]{0.95\linewidth}{$z^{o} \leftarrow$ Extract target task latent representation $f(\omega, \hat{\mathbf{s}}^{o})$}
\FOR{each source task $i = 1, \ldots, K$}
    \STATE \parbox[t]{0.95\linewidth}{$z^{i} \leftarrow$ Extract task i latent representation $f(\omega, \hat{\mathbf{s}}^i)$}
    \STATE \parbox[t]{0.95\linewidth}{$sim_i \leftarrow$ Substitute $z^{o}$ and $z^{i}$ into Eq.\eqref{equation_3} to compute the cosine similarity between tasks}
\ENDFOR
\STATE $\mathcal{S} \leftarrow \text{SelectTopSimilarTasks}(\mathcal{A}, \{sim_i\}_{i=1}^{K}, \gamma)$ 
\FOR{each source task $i \in \mathcal{S}$}
    \STATE \parbox[t]{0.95\linewidth}{\strut $w_i \leftarrow$ Substitute $sim_i$ into Eq.\eqref{equation_5} to compute the transfer weight of source task $i$ in the selected source subset $\mathcal{S}$ \strut}
\ENDFOR

{\fontsize{9}{12.5}\selectfont\texttt{// Stage 2: Solution-level transfer}}
\STATE \parbox[t]{0.95\linewidth}{$n_{opt} \leftarrow$ Total number $\lfloor \rho \cdot n_p \rfloor$ of injected solutions}
\STATE $\mathcal{P}_{inj} \leftarrow \emptyset$
\FOR{each task $i \in \mathcal{S}$}
    \STATE \parbox[t]{0.95\linewidth}{$n_i \leftarrow$ Number of injected solutions from source task $i$, computed as $\lfloor w_i \cdot n_{opt} \rfloor$}
    \STATE \parbox[t]{0.95\linewidth}{$\mathcal{P}_{sub} \leftarrow$ Select $n_i$ archived elite solutions from $\mathcal{D}^i$ using nondominated sorting and crowding-distance truncation}
    \STATE \parbox[t]{0.95\linewidth}{\strut $\mathcal{P}_{inj} \leftarrow \mathcal{P}_{inj} \cup \mathcal{P}_{sub}$\strut}
\ENDFOR
\STATE \parbox[t]{0.95\linewidth}{$\mathcal{P}_{rnd} \leftarrow$ Randomly sample $n_p - |\mathcal{P}_{inj}|$ solutions to fill the remaining population}
\STATE \parbox[t]{0.95\linewidth}{$\mathcal{P} \leftarrow \mathcal{P}_{inj} \cup \mathcal{P}_{rnd}$, initial population of the target task}

{\fontsize{9}{12.5}\selectfont\texttt{// Stage 3: Model-level transfer}}
\STATE $\mathcal{D}^{o} \leftarrow \phi$; \ $FE \leftarrow 0$
\WHILE{$FE < FE_{max}$}
    \IF{$|\mathcal{D}^{o}| > 0$}
        \STATE \parbox[t]{0.95\linewidth}{$G_{loc} \leftarrow$ Train a local GP surrogate using the evaluated target task samples $\mathcal{D}^{o}$}
        \STATE \parbox[t]{0.95\linewidth}{$G^{o} \leftarrow$ Substitute $\beta$, $\{w_i\}_{i\in\mathcal{S}}$, $\{G_i\}_{i\in\mathcal{S}}$, and $G_{loc}$ into Eq.\eqref{equation_6} to construct the ensemble surrogate of the target task}
    \ELSE
        \STATE \parbox[t]{0.95\linewidth}{$G^{o} \leftarrow$ target task ensemble surrogate by Eq.\eqref{equation_6} using only selected source task surrogates}
    \ENDIF
    
    \STATE \parbox[t]{0.95\linewidth}{$\mathcal{P}_{off} \leftarrow$ Generate offspring by evolution guided by the adaptive ensemble surrogate $G^{o}$}
    \STATE \parbox[t]{0.95\linewidth}{$\boldsymbol{\theta}_{new} \leftarrow$ Select promising candidate from $\mathcal{P}_{off}$ by the acquisition function}
    \STATE \parbox[t]{0.95\linewidth}{$\mathbf{F}_{new} \leftarrow$ High-fidelity WRF evaluation of $\boldsymbol{\theta}_{new}$}
    
    \STATE $\mathcal{D}^{o} \leftarrow \mathcal{D}^{o} \cup \{(\mathbf{\boldsymbol{\theta}_{new}}, \mathbf{F}_{new})\}$
    \STATE $FE \leftarrow FE + |\mathbf{\boldsymbol{\theta}_{new}}| $
    \STATE $\mathcal{P} \leftarrow \text{UpdatePopulation}(\mathcal{P} \cup \mathcal{P}_{off} \cup (\mathbf{\boldsymbol{\theta}_{new}}, \mathbf{F}_{new}))$ 
\ENDWHILE

{\fontsize{9}{12.5}\selectfont\texttt{// Stage 4: Archive update}}
\STATE \parbox[t]{0.95\linewidth}{$G^o \leftarrow$ $G^o$ trained from target task samples $\mathcal{D}^{o}$}
\STATE \parbox[t]{0.95\linewidth}{$\mathcal{A} \leftarrow$ Archive updated by storing $(\hat{\mathbf{s}}^{o}, \mathcal{D}^{o}, G^o)$}
\RETURN Non-dominated solution set $PS^{*}$ extracted from $\mathcal{D}^{o}$
\end{algorithmic}
\end{algorithm}

\noindent the Mean Squared Error (MSE) is adopted as the objective function:
\begin{equation}
  \label{equation_4}
  \begin{aligned}
  z_t &= \text{ResNet}(\text{Downsampling}_{\times n}(\hat{s}_t)) \\
  \hat{s}'_t &= \text{Upsampling}_{\times n}(\text{ResNet}(z_t)) \\
  \mathcal{L} &= \frac{1}{N} \sum_{i=1}^{N} \left\| \hat{s}_t^{(i)} - \hat{s}'_t{}^{(i)} \right\|_2^2
  \end{aligned}
\end{equation}
where $N$ represents the number of samples in a training batch, and $\|\cdot\|_2$ denotes the $L_2$ norm. 
By minimizing the loss function $\mathcal{L}$, the model compels the encoder to retain crucial feature information from the original data within the low-dimensional latent vector $z_t$, thereby achieving high-quality reconstruction.

\rev{Based on the cosine similarity between each source task and the target task, we first select \(\gamma\) similar source tasks from the task archive. This top-\(\gamma\) operation serves as a hard screening step that removes source tasks with clearly low relevance. After this subset has been determined, the pre-trained GP surrogate models associated with the selected source tasks are combined into an ensemble surrogate model. The ensemble weights are computed as}
\begin{equation}
  \label{equation_5}
  w_i = \frac{\exp(sim_i / \Gamma)}{\sum_{j=1}^{\revv{\gamma}} \exp(sim_j / \Gamma)}
\end{equation}
\rev{where \(\Gamma\) is a temperature coefficient. The resulting weights satisfy \(0<w_i<1\) and \(\sum_{i=1}^{\gamma}w_i=1\). This formula softly allocates the transfer contribution of each selected surrogate model according to its similarity to the target task. This weighting mechanism lies between a top-1 transfer strategy and an unweighted top-\(\gamma\) ensemble. When one selected source task is much more similar to the target task than the others, the ensemble assigns a larger weight to that source task. When several selected source tasks have comparable similarities, the ensemble can still exploit complementary knowledge from multiple sources. In this way, the weighted surrogate ensemble provides useful global guidance for the target task in the early stage of optimization, while reducing the risk of being dominated by a single misleading historical source task.}

\subsection{Adaptive Knowledge Transfer Mechanism}
Building upon the source task subset identified in the previous section, we propose an Adaptive Knowledge Transfer Mechanism. 
This mechanism reduces the influence of low-similarity source tasks while amplifying the impact of high-similarity ones. 
Facilitating knowledge transfer at both the model and solution levels, it provides a superior initialization for the optimization of newly arriving target tasks.

{\textbf{Model Level}}: We employ a weighted combination of surrogate models corresponding to multiple source tasks to approximate the current new task. 
The ensemble model assigns distinct weights to surrogate models from different source tasks, leveraging knowledge captured from different regions of the solution space by models trained in diverse environments. 
This constructs a more flexible approximation of the complex landscape of the new target task, which shares "partial similarities" with the source task subset. 
This mechanism serves as a buffer and error-correction strategy, enhancing the robustness of the transfer process. 
Furthermore, relying solely on an ensemble surrogate model constructed from historical tasks is insufficient to meticulously characterize the unique objective landscape of the new target task as the optimization progresses. 
Therefore, a local surrogate model is introduced to learn the specific information of the new target task, which is dynamically updated throughout the optimization.
Utilizing the ensemble weights derived in Section III-C, we construct an adaptive ensemble surrogate model for the current target task by aggregating the surrogate models $G_i^s$ corresponding to the selected subset of similar source tasks. In this work, GP regression is adopted as the underlying surrogate model. To capture the specific landscape of the target task, a local surrogate model, denoted as $G^{loc}$, is introduced. As the number of evaluated samples increases, $G^{loc}$ is iteratively updated throughout the optimization process. Consequently, the adaptive ensemble surrogate $G^o$ for the target task is formulated as:
\begin{equation}
\label{equation_6}
G^o = (1 - \beta) \sum_{i=1}^{\gamma} w_i \cdot G_i^s + \beta G^{loc}
\end{equation}
where $\beta = 1 - \exp(-c \cdot FE)$ represents a dynamic weight, and $\gamma$ denotes the number of selected source tasks. Here, $c$ regulates the rate of transition from being dominated by the ensemble surrogate model constructed from the source task subset to being dominated by the local surrogate model. 

{\textbf{Solution Level}}: During the optimization process of similar source calibration tasks, solution data evaluated by the NWP model are collected. 
We reuse the non-dominated solution sets from the collected data, injecting the optimal solution set into the new target task to facilitate a "warm start" of the optimization process. 
As historical calibration tasks, the optimal solution sets contain valuable information regarding potential regions in the solution space, which is highly beneficial for accelerating the exploration of the optimal solution set for the new target task. 
In the target task optimization, the population size is set to $n_p$. The initial population consists of two parts: the injected optimal solution set and a random solution set. 
The number of optimal solutions is $n_{opt}$, and the number of random solutions is $n_{ran} = n_p - n_{opt}$. This configuration achieves knowledge transfer while maintaining solution diversity.
Utilizing the ensemble model weights obtained in Section III-C, from the expensive evaluation solution set of each task in the similar source task subset, 
the number of solutions to be extracted is calculated proportional to the weights as $n_i =\lfloor w_i \cdot n_{opt} \rfloor$.
To ensure both the convergence and diversity of the injected solutions, we first perform non-dominated sorting on the expensive evaluation solution set of the $i$-th similar source task, partitioning it into distinct non-dominated levels $\mathcal{F}_1, \mathcal{F}_2, \dots, \mathcal{F}_l$. 
Based on the relationship between the size of the first non-dominated level $|\mathcal{F}_1|$ and the target extraction count $n_i$, the final screened optimal solution subset $o_i$ is defined as follows:
\begin{equation}
\label{equation_7}
\begin{cases} 
\mathcal{T}_{CD}(\mathcal{F}_1, n_i), & \text{if } |\mathcal{F}_1| > n_i \\
\mathcal{F}_1, & \text{if } |\mathcal{F}_1| = n_i \\
(\bigcup_{j=1}^{m-1} \mathcal{F}_j) \cup \mathcal{T}_{CD}(\mathcal{F}_m, n_{other}), & \text{if } |\mathcal{F}_1| < n_i 
\end{cases}
\end{equation}
where $\mathcal{T}_{CD}(\mathcal{F}, g)$ represents the crowding distance truncation operator, used to select the top $g$ optimal individuals from set $\mathcal{F}$ based on descending crowding distance. 
When $|\mathcal{F}_1| < n_i$, the algorithm sequentially absorbs solutions from subsequent levels until the $m$-th level cannot be fully included. At this point, the remaining number of required solutions is $n_{other} = n_i - \sum_{j=1}^{m-1} |\mathcal{F}_j|$. 
This strategy ensures that when the number of primary solutions is insufficient, secondary solutions are utilized for supplementation, while prioritizing the retention of non-dominated solutions with the widest distribution when truncating.

\section{Experiments}
\label{Section-IV}
\subsection{Experimental Settings}
The experimental evaluation of this study is structured into four primary components:
\begin{itemize}
  \item{\revv{A performance comparison of the proposed SEETO against four representative baseline algorithm under the constraint of a limited budget of expensive WRF evaluations, i.e., MOTPE \cite{akiba2019optuna}, MOASMO \cite{ref41}, KMO \cite{ref19}, and SAS-CKT-MO \cite{ref42}. \revvv{MOTPE is a multiobjective Bayesian optimization method for hyperparameter optimization.} MOASMO and KMO are state-of-the-art surrogate-assisted evolutionary algorithms specifically designed for WRF parameter calibration. SAS-CKT-MO is a general evolutionary transfer optimization baseline adapted from SAS-CKT \cite{ref42} for the current multiobjective WRF calibration scenario. To form SAS-CKT-MO, we expand the number of Gaussian Process Regression in SAS-CKT from one to two, to separately deal with the two objectives of temperature RMSE and wind speed RMSE. In addition, HV improvement is used in SAS-CKT-MO as the improvement criterion to guide the search, consistent with SEETO.}}
  \item{\revv{A two-part ablation study on the knowledge transfer mechanism is conducted. First, SEETO's knowledge transfer mechanism is separately incorporated into MOASMO and KMO, resulting in two variant algorithms, namely SEETO-MOASMO and SEETO-KMO, to verify the effectiveness of the knowledge transfer mechanism. Second, three simplified transfer variants, namely NTWS-SEETO, EIT-SEETO, and SET-SEETO, are constructed to analyze the contribution of different transfer components within SEETO.}}
  \item{A comparative analysis of the effectiveness of the employed meteorological representation extractor against two related meteorological state extraction algorithms, SCL \cite{ref43} and AtmoDist \cite{ref44}, within the context of WRF parameter calibration tasks.}
  \item{A parameter sensitivity analysis of SEETO to discuss the impact of parameter selection on optimization performance.}
\end{itemize}

\begin{table*}[!ht]
  \caption{Physical Schemes and Parameters Configuration}
  \label{tab:schemes_parameters}
  \centering
  \begin{tabularx}{\textwidth}{|
    >{\hsize=0.7\hsize\centering\arraybackslash}X|
    >{\hsize=1.0\hsize\centering\arraybackslash}X|
    c|c|c|
    >{\hsize=1.3\hsize\centering\arraybackslash}X|} 
  \hline
  \textbf{Physical process} & \textbf{Specific scheme} & \textbf{\makecell{Parameter}} & \textbf{Default} & \textbf{Range} & \textbf{Description} \\
  \hline
  Short-wave radiation & Dudhia shortwave radiation scheme & \textbf{cssca} & 0.00001 & $[5e^{-6}\ , 2e^{-5}]$ & Scattering tuning parameter($m^2 kg^{-1}$) \\
  \hline
  Long-wave radiation & RRTM longwave radiation scheme & --- & --- & --- & --- \\
  \hline
  Atmospheric boundary layer & Monin-Obukhov surface layer scheme & --- & --- & --- & --- \\
  \hline
  Land surface & unified Noahland-surface model & \textbf{porsl} & 1 & [0.5,2] & Multiplier for the saturated soil water content \\
  \hline
  Cumulus & Kain-Fritsch Eta cumulus scheme & --- & --- & --- & --- \\
  \hline
  Planetary boundary layer & Yonsei University planetary boundary layer scheme & \textbf{pfac} & 2 & [1,3] & Profile shape exponent used to calculate the momentum diffusivity coefficient \\
  \hline
  \multirow{2}{=}{\centering Microphysics} & 
  \multirow{2}{=}{\centering WSM six-class Graupel microphysics scheme} & 
  \textbf{\makecell{ice\_stok\\es\_fac}} & 14900 & [8000,30000] & Scaling factor applied to ice fall velocity($s^{-1}$) \\
  \cline{3-6}
   & & \textbf{dimax} & 0.0005 & $[3e^{-4}\ , 8e^{-4}]$ & Limiting maximum value for the cloud-ice diameter(m) \\
  \hline
  \end{tabularx}
\end{table*}

To conduct our experiments, the following experimental setup was adopted for the WRF parameter calibration tasks:

\textbf{Calibration Scenario}: The spatial domain for the WRF forecast covers the longitude and latitude range of $38.0^\circ N - 41.75^\circ N$ and $114.5^\circ E - 118.25^\circ E$. \revvv{The physical parameterization schemes adopted for the WRF model and the parameters selected for calibration are consistent with those in \cite{ref19}, as detailed in Table~\ref{tab:schemes_parameters}. The symbol “---” indicates that the parameters of the corresponding physical scheme are held fixed. These schemes are listed solely to provide a complete description of the combination of physical parameterization schemes used.} RMSE is employed as the error metric in the calibration tasks:
\begin{equation}
\label{equation_9}
RMSE = \sqrt{\frac{1}{N \times T}\sum_{i=1}^{N}\sum_{t=1}^{T}(s_{i,t} - \hat{s}_{i,t})^2}
\end{equation}
where $s_{i,t}$ and $\hat{s}_{i,t}$ represent the simulated value and the observed value, respectively, for the $i$-th grid cell at time step $t$, and $N$ denotes the total number of grid cells.

\textbf{Data Sources}: Meteorological data were obtained from the ERA5 \footnote{\revv{The fifth generation ECMWF atmospheric reanalysis of the global climate, where ECMWF stands for the European Centre for Medium-Range Weather Forecasts.}} reanalysis dataset \cite{ref45} and the China Meteorological Administration Land Data Assimilation System (CLDAS-V2.0) \footnote{https://data.cma.cn/data/cdcdetail/dataCode/NAFP\_CLDAS2.0\_RT.html} real-time product dataset. 
ERA5 data serve as the meteorological field input for WRF, providing the necessary data to construct the meteorological states.
Since the forecast results output by WRF possess a higher resolution compared to the ERA5 data, the CLDAS dataset is selected as the calibration target to facilitate calibration at a high precision resolution.

\textbf{Task Sets}: To provide a transferable source task set for SEETO, we partitioned the period from 2017/07/01 00:00 to 2017/07/11 00:00 into $K=20$ source calibration tasks at 12-hour intervals. 
The target task set consists of 10 tasks selected from August 2017 (08/03 11:00-08/03 23:00, 08/06 09:00-08/06 21:00, 08/08 14:00-08/09 02:00, 08/12 09:00-08/12 21:00, 08/15 21:00-08/16 09:00, 08/18 15:00-08/19 03:00, 08/21 09:00-08/21 21:00, 08/24 09:00-08/24 21:00, 08/27 20:00-08/28 08:00, 08/29 18:00-08/30 06:00). 
These are denoted sequentially as "task 1" through "task 10," sorted chronologically, with an interval of approximately 3 days between tasks. \revv{Within the source task set, target tasks 1–7 can identify highly similar source tasks, \revvv{whereas no highly similar source tasks are available for target tasks 8–10.} This is consistent with realistic calibration scenarios that not all tasks are similar.}

\textbf{Expensive Evaluation Budget}: \revvv{According to the operational workflow described in Section~II-A, parameter calibration is only one component of a 12-hour rolling forecast cycle, and time must also be reserved for subsequent procedures such as data preprocessing, data assimilation, numerical integration, forecast product analysis, and release. As a result, the practical time available for WRF parameter calibration is usually only about 1--1.5 hours. Under the current experimental environment, a single WRF run takes 31min 8s on a single CPU core and 4min 26s with 32 CPU cores in parallel. Therefore, 20 evaluations require 1h 28min 40s under the 32-core setting, which is broadly consistent with operational time constraints.} 

\revvv{For this reason, we use $\mathrm{FE}=20$ as the primary reporting point, since it more closely reflects realistic scenarios in which parameter decisions must be made within a limited time. On the other hand, the parallel speedup of WRF is clearly sublinear, making it difficult to increase the number of executable evaluations proportionally by continuously adding computational resources. Meanwhile, in practical multi-region operational scenarios, it is also unrealistic to allocate large-scale parallel resources to a single calibration task for a prolonged period. Therefore, we further set $\mathrm{FE}=60$ as an extended evaluation budget to examine the performance evolution of different methods over a longer optimization horizon.}

\textbf{Hardware and software environment}: An AMD Ryzen Threadripper PRO 7985WX processor with 64 cores and 128GB of RAM, the GPU model is an NVIDIA RTX 5880 Ada Generation with 48GB of VRM. 
The NWP model employed is the WRF model version 4.5. The optimization algorithms involved in this study were implemented based on the PlatEMO platform \cite{ref47}.

\begin{table}[!h]
  \caption{Parameter Settings for SEETO\label{tab:seeto_parameters}}
  \centering
  \begin{tabularx}{\columnwidth}{|c|c|>{\centering\arraybackslash}X|c|} 
  \hline
  \textbf{Module} & \textbf{Parameter} & \textbf{Description} & \textbf{Value} \\ 
  \hline
  Feature Extractor & $\lambda$ & Number of meteorological state elements & 19 \\
  \hline
  \multirow{6}{*}{Ensemble Surrogate} & $\gamma$ & Number of selected tasks & 5 \\
  \cline{2-4}
   & $\Gamma$ & Temperature coefficient & 0.065 \\
  \cline{2-4}
   & $c$ & Dynamic weight $\beta$ control parameter & \begin{tabular}{@{}c@{}}0.038\\ or\\ 0.017\end{tabular} \\
  \cline{2-4}
   & $\tau$ & Task similarity threshold & 0.7 \\
  \hline
  Solution Injection & $\rho$ & Ratio of injected solutions & 0.2 \\
  \hline
  WRF Evaluation & $r$ & Reference point & $\left(1,3\right)$ \\
  \hline
  \end{tabularx}
\end{table}

\begin{table}[t]
\caption{\revv{The Structure Hyper-parameters of the Meteorological State Representation Extractor}}
\label{tab:autoencoder_hyperparameters}
\centering
\renewcommand{\arraystretch}{1.2}
\setlength{\tabcolsep}{4pt}
\begin{tabular}{>{\centering\arraybackslash}m{0.28\columnwidth} >{\centering\arraybackslash}m{0.65\columnwidth}}
\toprule
\textbf{Model Module} & \textbf{Structure Hyper-parameter} \\
\midrule
Encoder Structure 
& 3 stages (\texttt{ConvNormAct + ResidualBlock}), with channels [64, 128, 256] \\

Decoder Structure 
& 3 stages (\texttt{ResidualBlock + ConvNormActT}), with channels [256, 128, 64] \\

Output Layer 
& \texttt{Conv2d} with 19 output channels \\

Latent Dimension 
& $256$ \\

ConvNormAct 
& \texttt{Conv2d}, normalization, and activation \\

ConvNormActTrans 
& \texttt{ConvTranspose2d}, normalization, and activation \\

Normalization 
& GroupNorm \\

Activation Function 
& GELU \\

Optimizer 
& AdamW with $lr = 8 \times 10^{-4}$ and $weight\_decay = 10^{-5}$ \\
\bottomrule
\end{tabular}
\end{table}

\subsection{Algorithm Settings}
The algorithmic settings required for SEETO are presented in Table~\ref{tab:seeto_parameters}, primarily comprising the following aspects:

\textbf{Meteorological State Representation Extraction}: The meteorological state is composed of multiple meteorological variables, with the selection referencing WeatherBench 2 \cite{ref48}. 
This includes 4 single-level variables (2m temperature, 10m u-component of wind, 10m v-component of wind, and mean sea level pressure) and 5 upper-air variables (geopotential, temperature, u-component of wind, v-component of wind, and relative humidity). 
 While the original dataset distributes each upper-air variable across 13 pressure levels, the number of input variables influences both the accuracy of representation extraction and training costs. 
To balance accuracy with computational efficiency, the number of meteorological variables is set to $\lambda = 19$ (comprising 5 upper-air variables across 3 distinct pressure levels and 4 single-level variables). \revv{The meteorological state representation extractor adopts an encoder-decoder architecture, with the detailed network structure and hyperparameter settings presented in Table~\ref{tab:autoencoder_hyperparameters}.} \revvv{The model was trained on data from 2013 to 2016.}

\textbf{Ensemble Surrogate Construction}: The similarity between different target tasks and each source task varies. While high-similarity tasks facilitate transfer, transferring from low-similarity source tasks often leads to "negative transfer." Consequently, the top $\gamma = 5$ similar source tasks from the archive are selected as the source task subset for knowledge transfer. Additionally, the temperature coefficient for similarity normalization is set to $\Gamma = 0.065$, concentrating the ensemble surrogate weights on the most similar source tasks. The control parameter $c$ for the dynamic weight $\beta$, which governs the integration of the local surrogate into the ensemble, is set to $0.038$ or $0.017$. The specific value depends on whether high-similarity tasks exist within the source subset ($w_i \ge \tau$) or not ($w_i < \tau$), where the threshold is set to $\tau = 0.7$.

\textbf{Optimal Solution Set Injection}: The population size is set to $n_p = 100$. In the initial population, the proportion of reused optimal solutions is set to $\rho = 0.2$. Specifically, the initial population consists of $n_{opt} = \lfloor \rho \cdot n_p \rfloor$ optimal individuals reused from past environments, while the remaining $n_{ran} = n_p - n_{opt}$ individuals are randomly generated.

\textbf{\revv{Performance Evaluation Metrics}}:\revv{Each instance was independently run 10 times. For the bi-objective calibration problem of minimizing wind speed RMSE and temperature RMSE, Hypervolume (HV) was used as the primary metric and is reported as mean $\pm$ standard deviation. HV was computed on the nondominated set in the two-dimensional objective space. The reference point was defined as $r=(r_{\text{wind}},r_{\text{temp}})=\left(\max_{x\in\mathcal{P}} f_1(x)+\epsilon_1,\ \max_{x\in\mathcal{P}} f_2(x)+\epsilon_2\right)$,where $\mathcal{P}$ denotes the combined nondominated set obtained from all methods and runs. In practice, the worst value on each objective was multiplied by 1.1 and got rounded to determine the corresponding component of the reference point, resulting in $r=(1,3)$.}

\begin{table*}[!t]
  \centering
  \caption{\revv{Comparison of HV Performance and Search Efficiency}}
  \label{tab:hv20comparison}
  \renewcommand{\arraystretch}{0.95}
  \setlength{\tabcolsep}{3pt}
  \scriptsize

  \resizebox{\textwidth}{!}{%
  \begin{tabular}{@{}ccccccccccccccc@{}}
    \toprule

    \multirow{2}{*}{Problem} &
    \multirow{2}{*}{\makecell{WRF\\FE}} &
    \multicolumn{3}{c}{\revvv{MOTPE}} &
    \multicolumn{3}{c}{MOASMO} &
    \multicolumn{3}{c}{KMO} &
    \multicolumn{3}{c}{SAS-CKT-MO} &
    SEETO \\

    \cmidrule(lr){3-5}
    \cmidrule(lr){6-8}
    \cmidrule(lr){9-11}
    \cmidrule(lr){12-14}
    \cmidrule(lr){15-15}

    & &
    HV & \makecell{$\Delta$HV\\(\%)} & \makecell{Add.FE\\(\%)}
    & HV & \makecell{$\Delta$HV\\(\%)} & \makecell{Add.FE\\(\%)}
    & HV & \makecell{$\Delta$HV\\(\%)} & \makecell{Add.FE\\(\%)}
    & HV & \makecell{$\Delta$HV\\(\%)} & \makecell{Add.FE\\(\%)}
    & HV \\

    \midrule

    \cellcolor{mygreen}task 1 & 20
    & \HVmeanstd{2.0473E-01}{7.78E-03} & -6.86\% & $>200\%$
    & \HVmeanstd{2.0655E-01}{8.17E-03} & -6.41\% & 45.00\%
    & \HVmeanstd{2.0755E-01}{7.73E-03} & -5.90\% & 30.00\%
    & \HVmeanstd{2.0640E-01}{5.16E-03} & -6.49\% & 55.00\%
    & \HVbestmeanstd{2.1980E-01}{9.17E-04}
      {\P\dagger\ddagger\S}
    \\

    \cellcolor{mygreen}task 2 & 20
    & \HVmeanstd{1.3651E-01}{4.59E-03} & -9.30\% & $>200\%$
    & \HVmeanstd{1.4097E-01}{6.49E-03} & -6.77\% & 105.00\%
    & \HVmeanstd{1.4297E-01}{7.12E-03} & -5.27\% & 35.00\%
    & \HVmeanstd{1.4366E-01}{7.17E-04} & -4.77\% & 25.00\%
    & \HVbestmeanstd{1.5051E-01}{8.79E-04}
      {\P\dagger\ddagger\S}
    \\

    \cellcolor{mygreen}task 3 & 20
    & \HVmeanstd{6.1190E-01}{7.69E-03} & -1.48\% & 165.00\%
    & \HVmeanstd{6.1215E-01}{7.34E-03} & -1.46\% & 30.00\%
    & \HVmeanstd{6.1235E-01}{7.90E-03} & -1.43\% & 5.00\%
    & \HVmeanstd{6.1451E-01}{8.43E-03} & -1.07\% & 10.00\%
    & \HVbestmeanstd{6.2110E-01}{7.33E-04}
      {\P\dagger\ddagger\S}
    \\

    \cellcolor{mygreen}task 4 & 20
    & \HVmeanstd{8.7648E-02}{6.18E-03} & -16.38\% & $>200\%$
    & \HVmeanstd{8.1330E-02}{6.71E-03} & -28.88\% & $>200\%$
    & \HVmeanstd{8.1380E-02}{6.82E-03} & -28.80\% & 30.00\%
    & \HVmeanstd{9.6413E-02}{9.22E-04} & -8.72\% & 55.00\%
    & \HVbestmeanstd{1.0482E-01}{7.05E-04}
      {\P\dagger\ddagger\S}
    \\

    \cellcolor{mygreen}task 5 & 20
    & \HVmeanstd{5.2034E-01}{8.63E-04} & -0.19\% & 110.00\%
    & \HVmeanstd{5.1515E-01}{6.52E-03} & -1.20\% & 80.00\%
    & \HVmeanstd{5.1585E-01}{6.16E-03} & -1.06\% & 25.00\%
    & \HVmeanstd{5.1817E-01}{8.83E-04} & -0.61\% & 30.00\%
    & \HVbestmeanstd{5.2134E-01}{3.33E-04}
      {\P\dagger\ddagger}
    \\

    \cellcolor{mygreen}task 6 & 20
    & \HVmeanstd{6.0602E-01}{6.38E-03} & -1.12\% & 20.00\%
    & \HVmeanstd{6.0025E-01}{7.21E-03} & -2.11\% & 25.00\%
    & \HVmeanstd{6.0132E-01}{7.17E-03} & -1.94\% & 30.00\%
    & \HVmeanstd{5.9895E-01}{7.18E-04} & -2.39\% & 25.00\%
    & \HVbestmeanstd{6.1289E-01}{6.46E-04}
      {\P\dagger\ddagger\S}
    \\

    \cellcolor{mygreen}task 7 & 20
    & \HVmeanstd{3.4011E-01}{7.32E-04} & -5.88\% & $>200\%$
    & \HVmeanstd{3.4125E-01}{8.21E-03} & -5.89\% & 45.00\%
    & \HVmeanstd{3.4146E-01}{8.29E-03} & -5.83\% & 50.00\%
    & \HVmeanstd{3.4023E-01}{8.04E-04} & -6.21\% & 50.00\%
    & \HVbestmeanstd{3.6136E-01}{9.95E-04}
      {\P\dagger\ddagger\S}
    \\

    \midrule

    \cellcolor{mygray}task 8 & 20
    & \HVmeanstd{5.0376E-01}{6.86E-03} & -1.28\% & 100.00\%
    & \HVmeanstd{5.0341E-01}{7.16E-03} & -1.35\% & 50.00\%
    & \HVmeanstd{5.0383E-01}{7.29E-03} & -1.26\% & 50.00\%
    & \HVmeanstd{5.0245E-01}{7.13E-04} & -1.54\% & 40.00\%
    & \HVbestmeanstd{5.1019E-01}{7.24E-04}
      {\P\dagger\ddagger\S}
    \\

    \cellcolor{mygray}task 9 & 20
    & \HVmeanstd{2.7167E-01}{5.29E-04} & -1.50\% & 80.00\%
    & \HVmeanstd{2.7231E-01}{5.41E-03} & -1.25\% & 25.00\%
    & \HVmeanstd{2.7211E-01}{5.31E-03} & -1.33\% & 20.00\%
    & \HVmeanstd{2.7027E-01}{7.15E-04} & -2.02\% & 25.00\%
    & \HVbestmeanstd{2.7572E-01}{8.94E-05}
      {\P\dagger\ddagger\S}
    \\

    \cellcolor{mygray}task 10 & 20
    & \HVmeanstd{5.2341E-01}{6.75E-04} & -0.22\% & 50.00\%
    & \HVmeanstd{5.2157E-01}{6.40E-03} & -0.57\% & 35.00\%
    & \HVmeanstd{5.2177E-01}{6.45E-03} & -0.53\% & 5.00\%
    & \HVmeanstd{5.1811E-01}{7.05E-04} & -1.24\% & 55.00\%
    & \HVbestmeanstd{5.2456E-01}{8.02E-04}
      {\S}
    \\

    \midrule

    \multirow{2}{*}{Average Run Time}
    & 20
    & \multicolumn{3}{c}{1h 15m 16s}
    & \multicolumn{3}{c}{1h 27min 36s}
    & \multicolumn{3}{c}{1h 41min 15s}
    & \multicolumn{3}{c}{1h 31min 23s}
    & 1h 28min 49s
    \\

    & 60
    & \multicolumn{3}{c}{3h 50m 42s}
    & \multicolumn{3}{c}{4h 45min 31s}
    & \multicolumn{3}{c}{5h 10min 20s}
    & \multicolumn{3}{c}{4h 58min 34s}
    & 4h 51min 40s
    \\

    \bottomrule
  \end{tabular}%
  }
\end{table*}

\begin{figure*}[!ht]
\centering
\setlength{\fboxsep}{0.5pt} 
\newcommand{\myhspace}{\hspace{1pt}}

\subfloat{\label{fig:task_1}\colorbox{mygreen}{\includegraphics[width=0.195\textwidth]{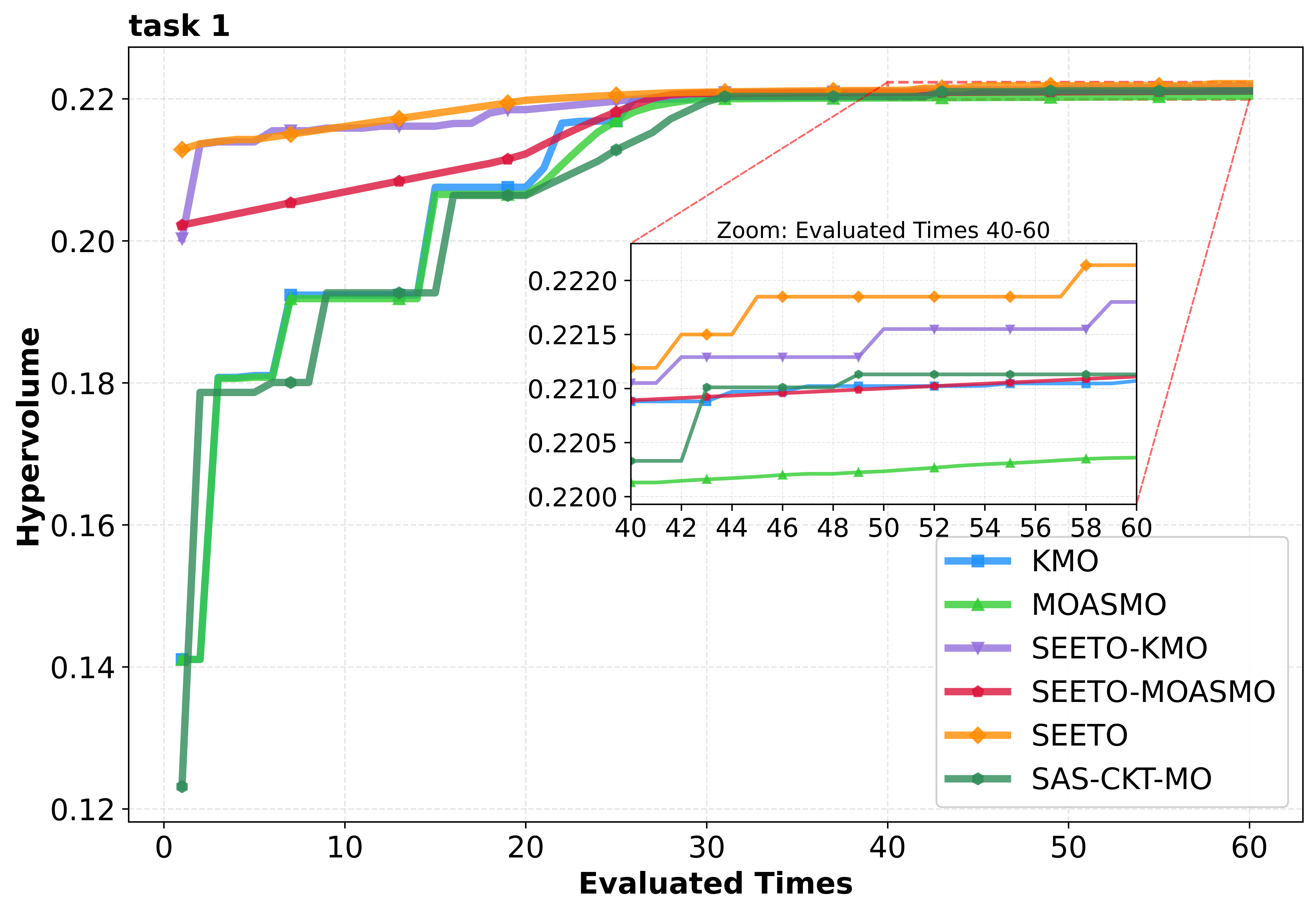}}}\myhspace
\subfloat{\label{fig:task_2}\colorbox{mygreen}{\includegraphics[width=0.195\textwidth]{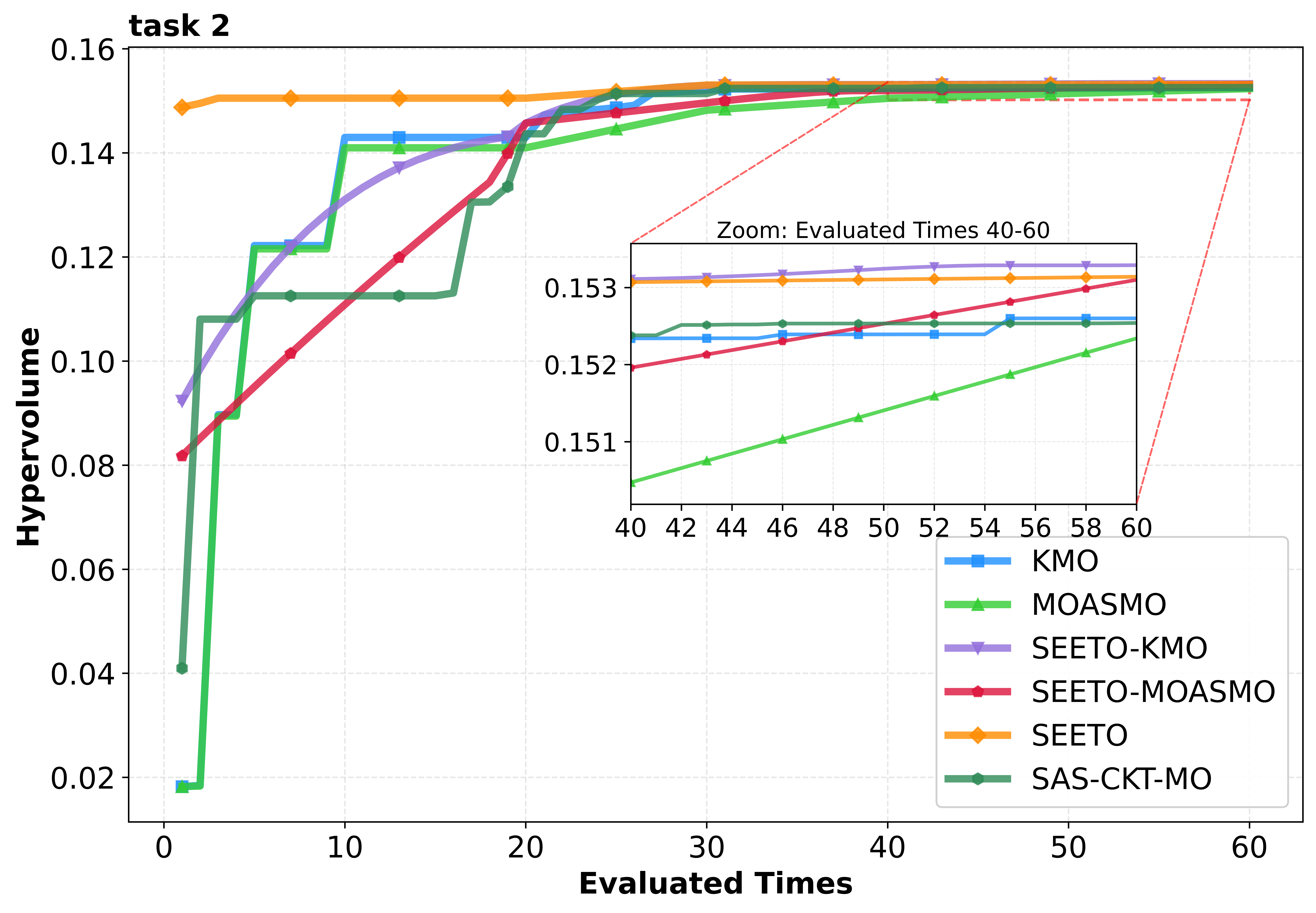}}}\myhspace
\subfloat{\label{fig:task_3}\colorbox{mygreen}{\includegraphics[width=0.195\textwidth]{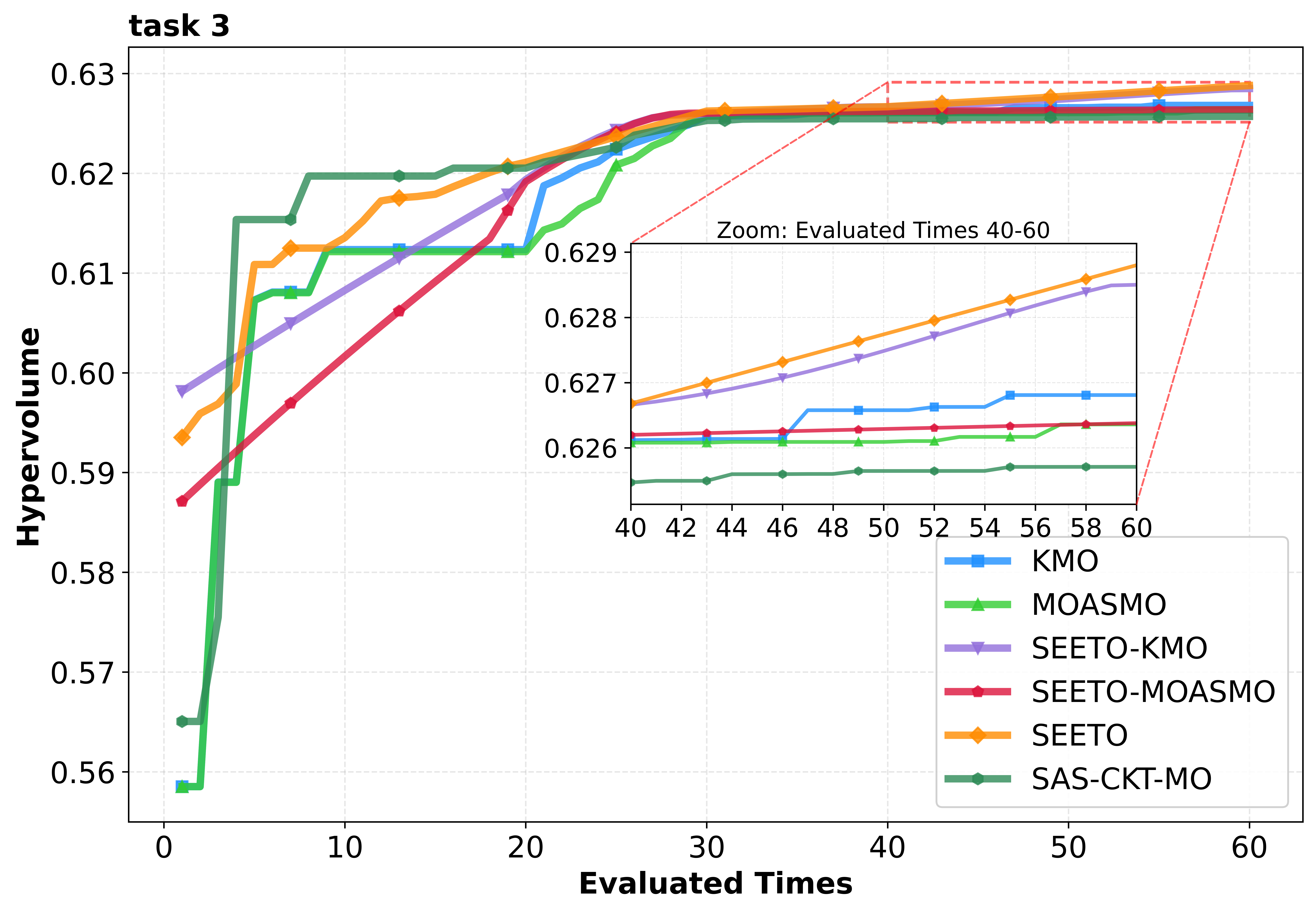}}}\myhspace
\subfloat{\label{fig:task_4}\colorbox{mygreen}{\includegraphics[width=0.195\textwidth]{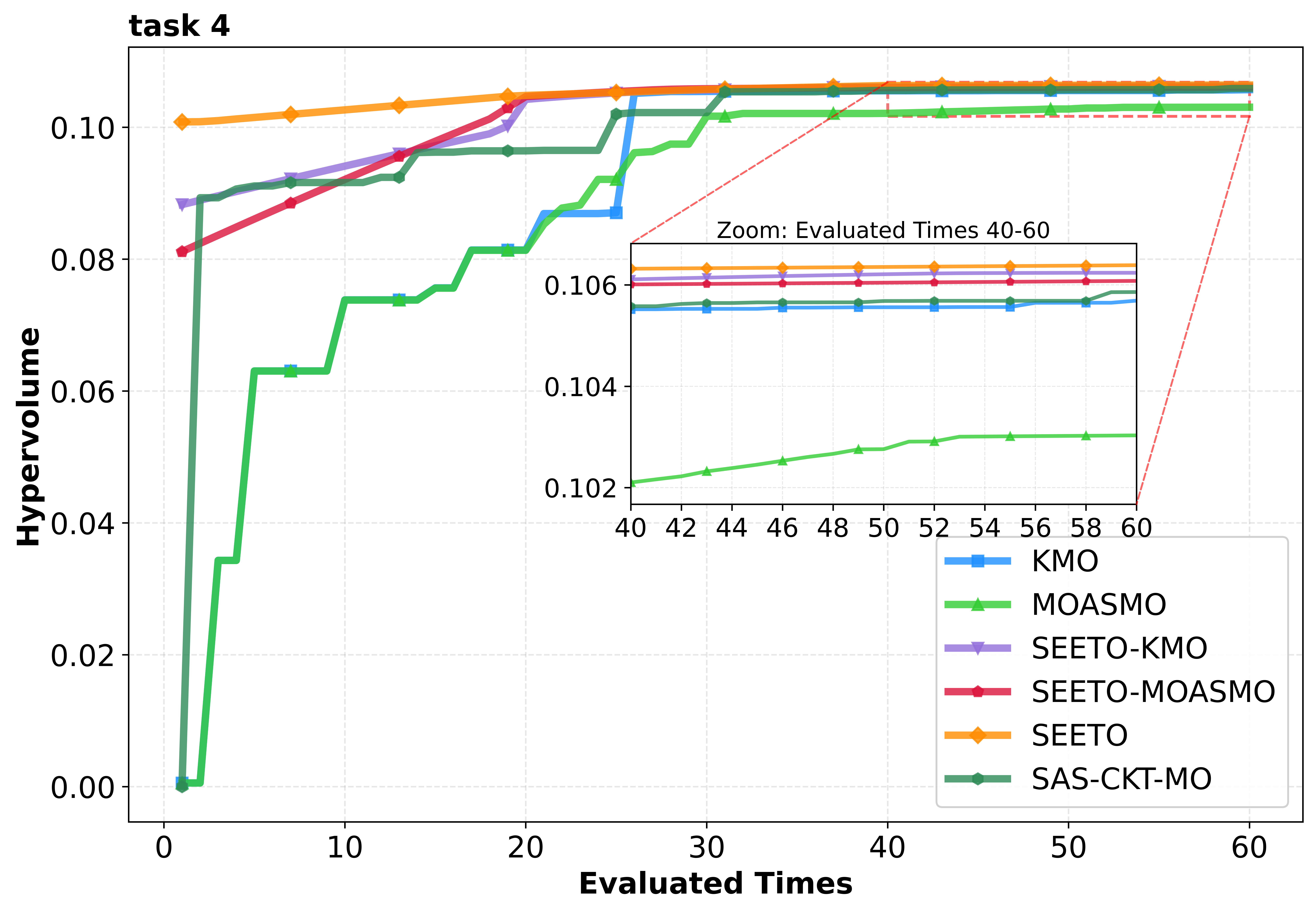}}}\myhspace
\subfloat{\label{fig:task_5}\colorbox{mygreen}{\includegraphics[width=0.195\textwidth]{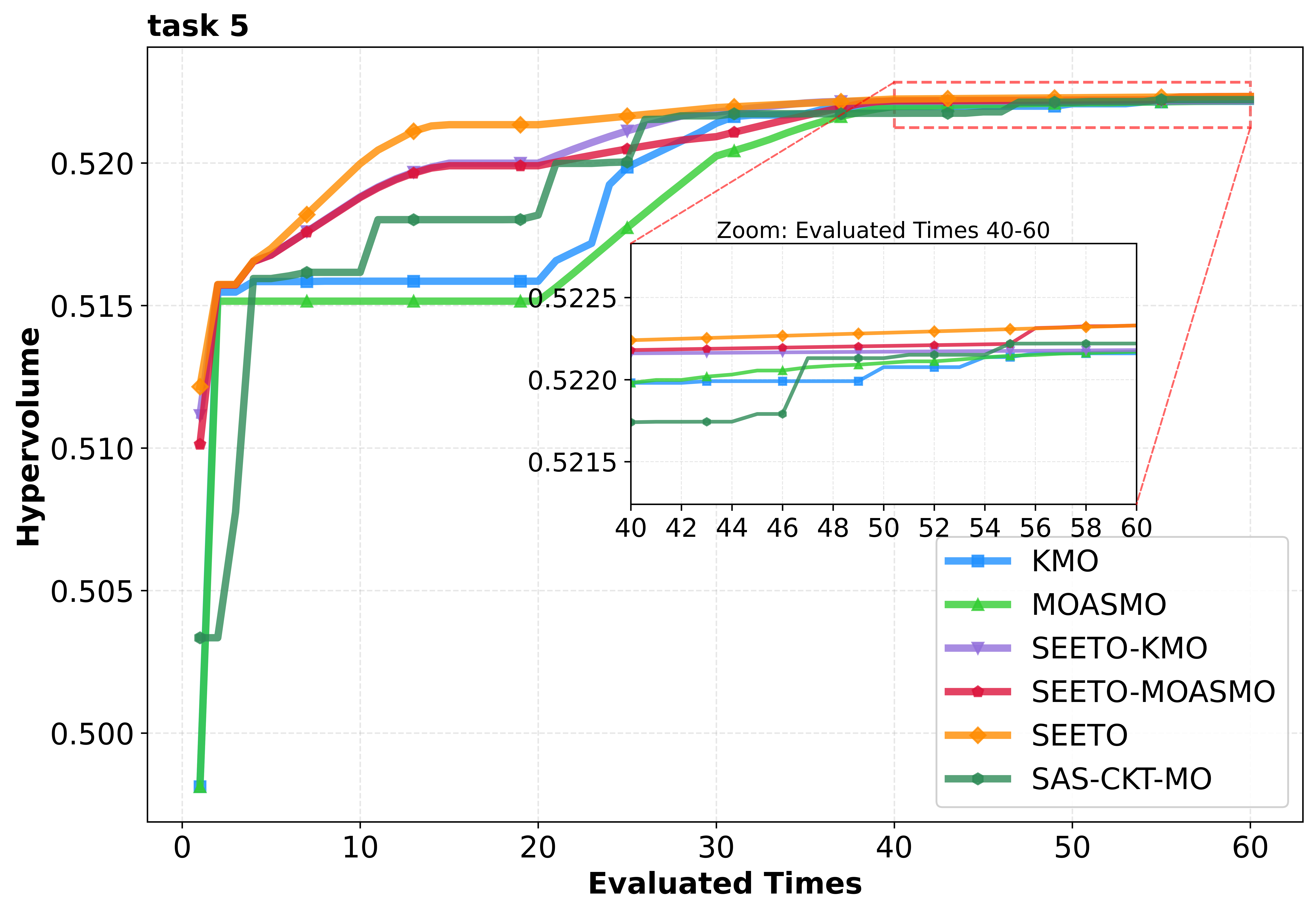}}}

\vspace{-8pt} 

\subfloat{\label{fig:task_6}\colorbox{mygreen}{\includegraphics[width=0.195\textwidth]{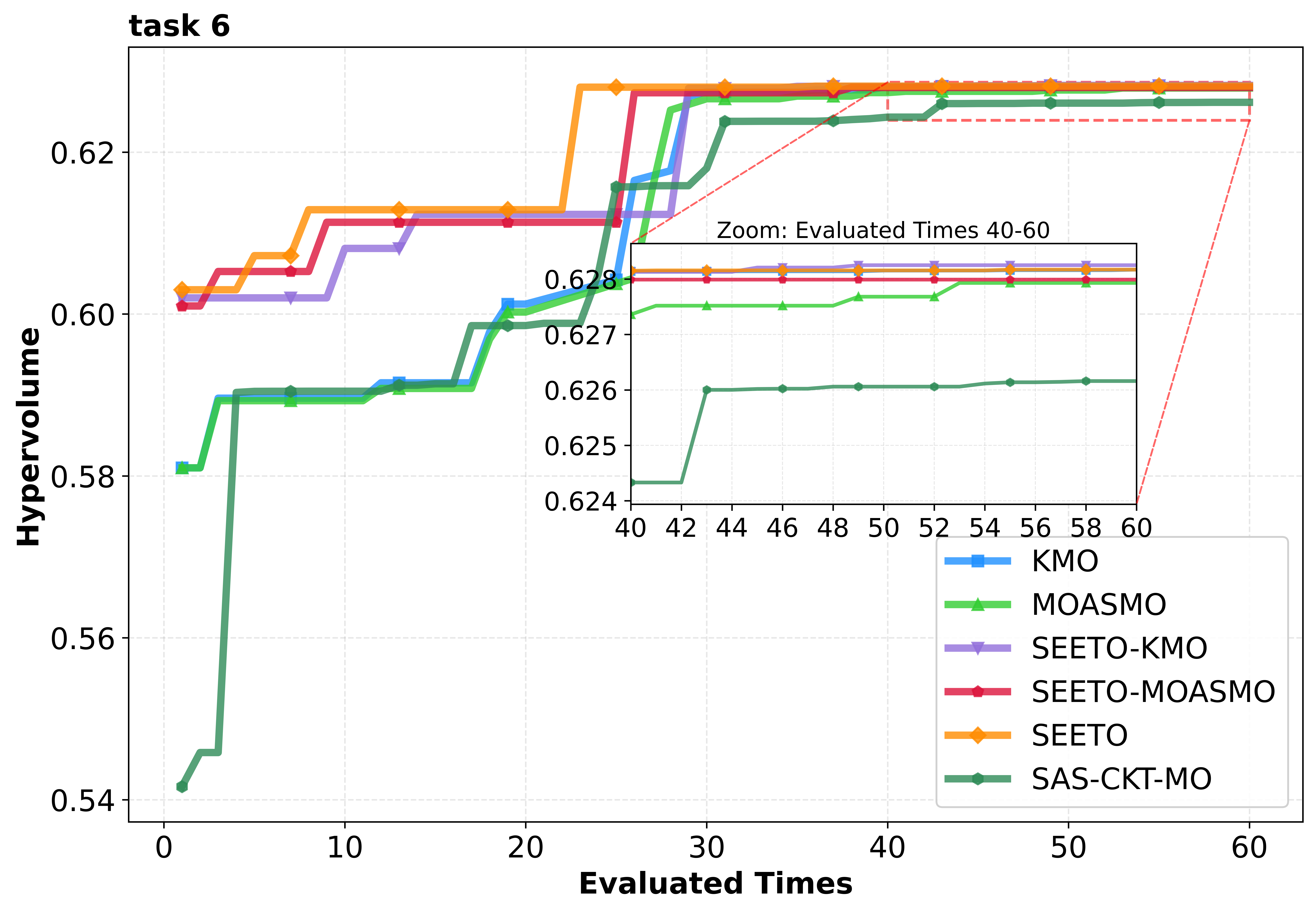}}}\myhspace
\subfloat{\label{fig:task_7}\colorbox{mygreen}{\includegraphics[width=0.195\textwidth]{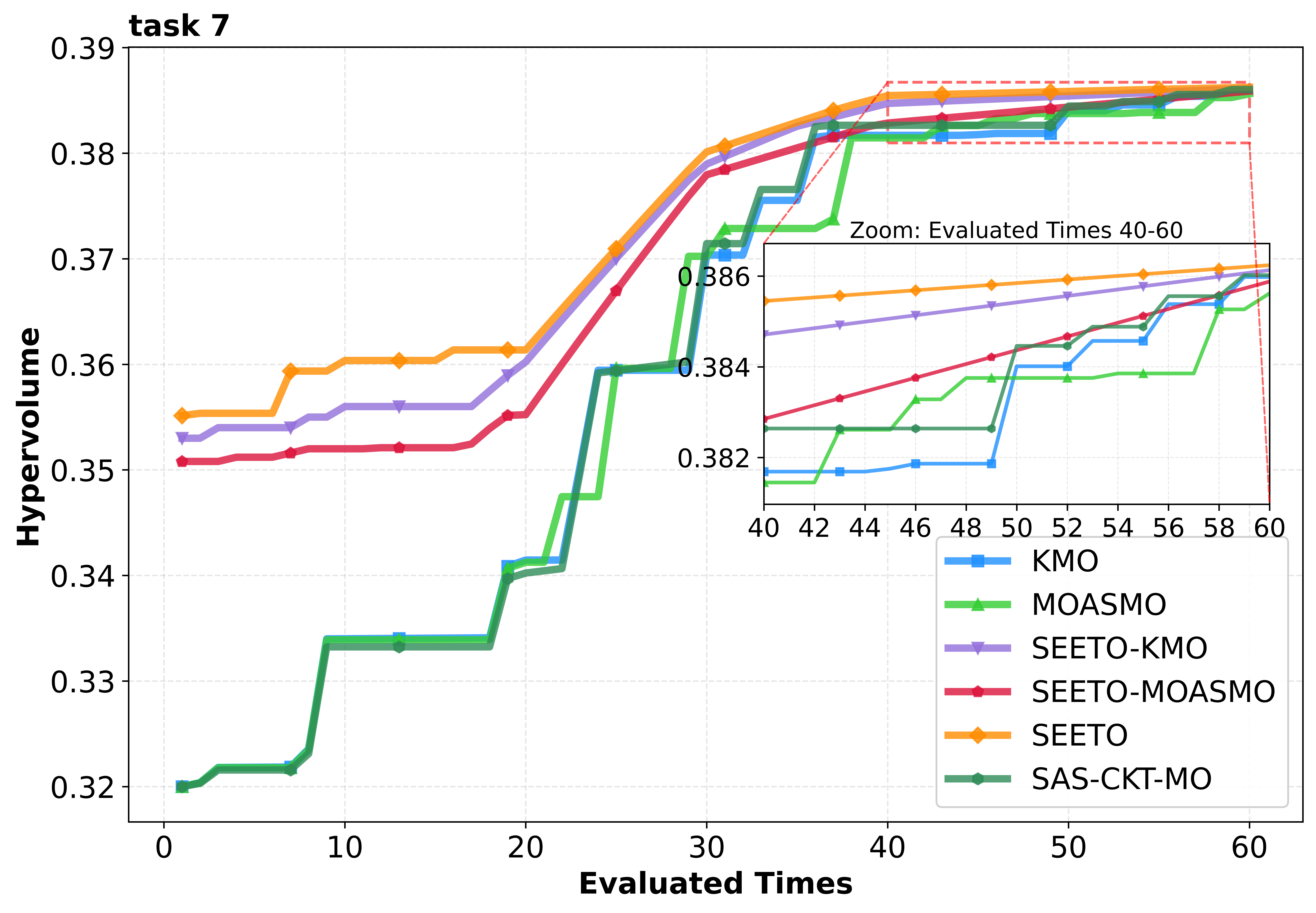}}}\myhspace
\subfloat{\label{fig:task_8}\colorbox{mygray}{\includegraphics[width=0.195\textwidth]{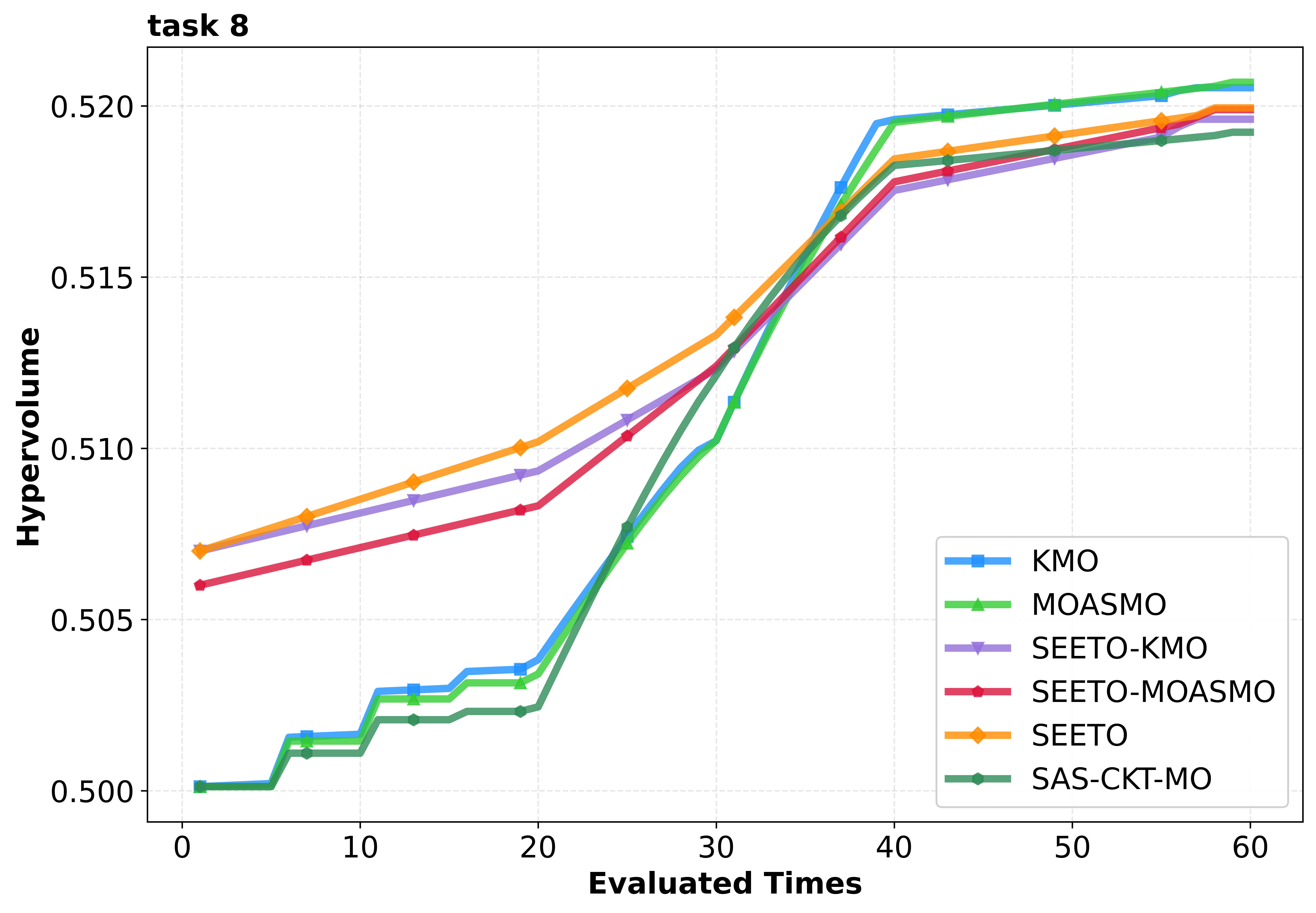}}}\myhspace
\subfloat{\label{fig:task_9}\colorbox{mygray}{\includegraphics[width=0.195\textwidth]{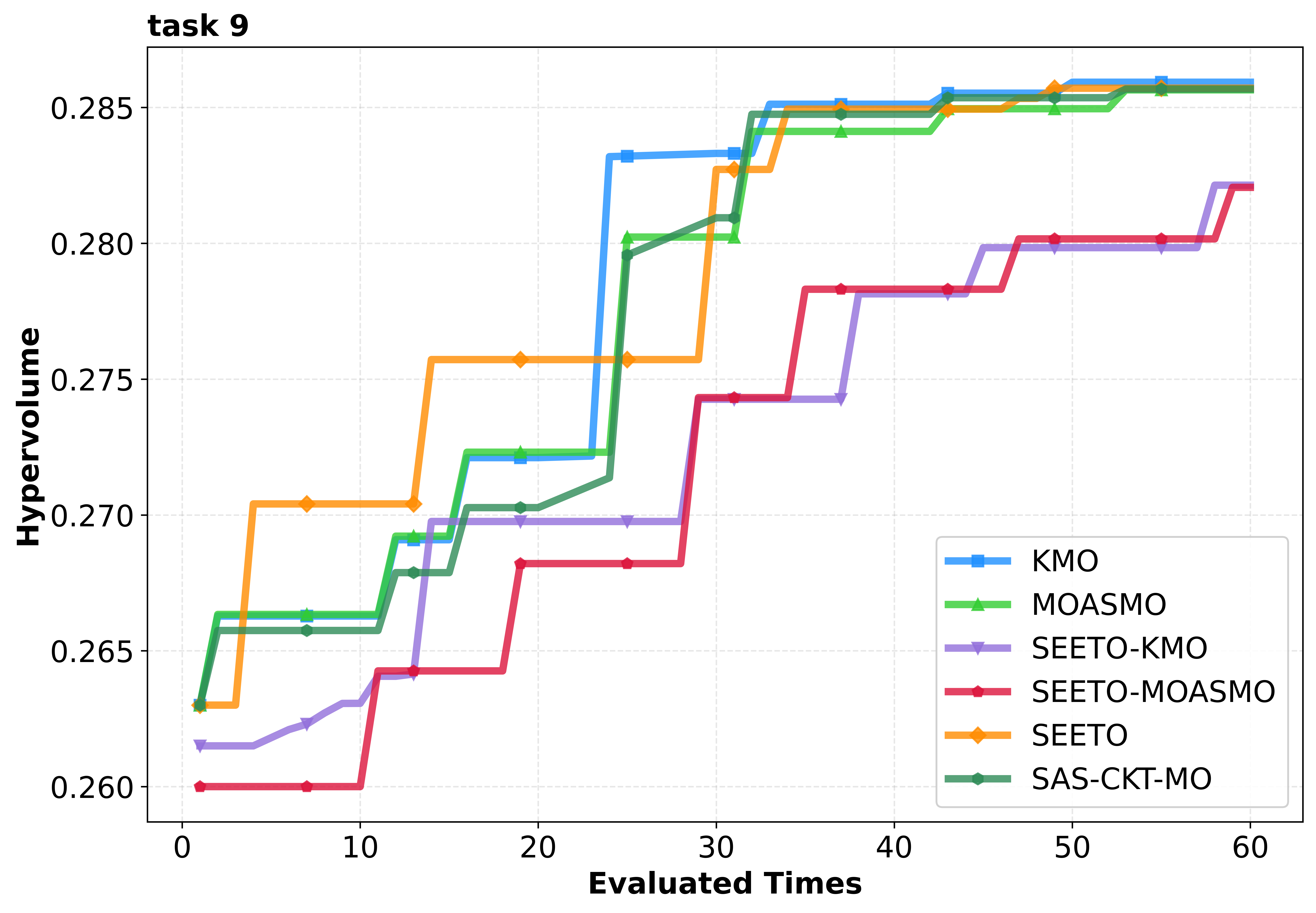}}}\myhspace
\subfloat{\label{fig:task_10}\colorbox{mygray}{\includegraphics[width=0.195\textwidth]{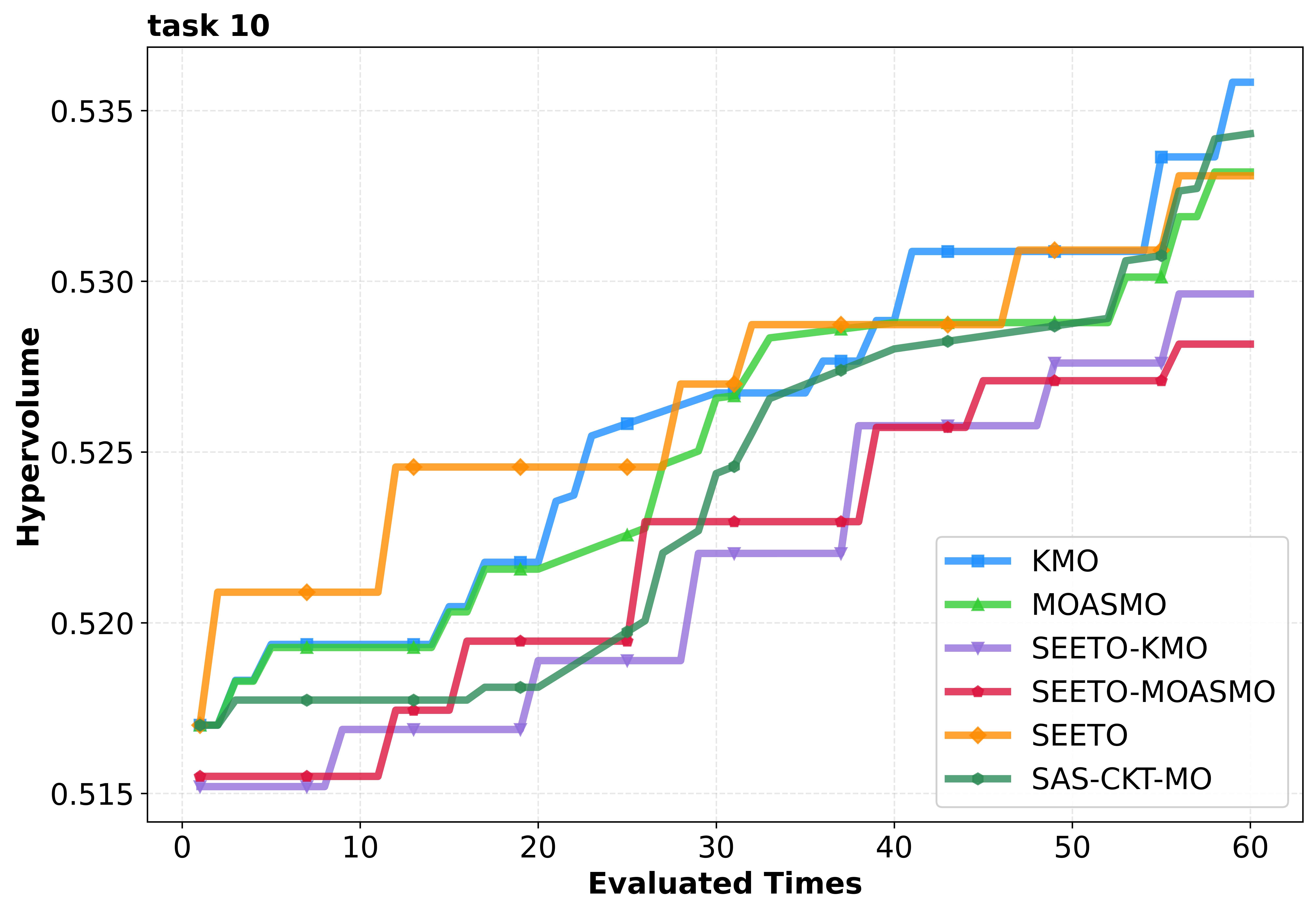}}}

\caption{\revv{HV evolution trajectories across 60 WRF evaluations. The subfigures with} \colorbox{mygreen}{green} \revv{background correspond to \revvv{Tasks 1--7 (High-similarity tasks)}, and those with} \colorbox{mygray}{gray} \revv{background correspond to \revvv{Tasks 8--10 (Low-similarity tasks)}.} Fig.~S1 in the supplementary material presents a full-size, high-resolution version of the experimental results, showing the task-level details more clearly.}
\label{fig:all_task_hv_line_colored}
\end{figure*}

\subsection{Performance Comparison of Calibration Algorithms}
\revv{\revvv{In this subsection, we compare SEETO with two state-of-the-art algorithms for single WRF calibration tasks, namely KMO and MOASMO, as well as with SAS-CKT-MO, a recent general evolutionary transfer optimization algorithm, and MOTPE, a non-evolutionary multi-objective Bayesian optimization method based on probabilistic density estimation.} During the initialization phase, both KMO and MOASMO employ Quasi-Monte Carlo (QMC) sampling to generate 20 parameter vectors, which are then evaluated by WRF to construct the surrogate models for the problem.}

The comparative results are presented in Table~\ref{tab:hv20comparison}. Here, \revv{$\Delta HV(\%) = \frac{HV_{baselines}^{(20)} -  HV_{\mathrm{SEETO}}^{(20)}}{HV_{\mathrm{SEETO}}^{(20)}} \times 100\%$} denotes the percentage difference in HV values between the baselines and SEETO at the $20$-th evaluation. 
\revv{$Add. FE(\%) = \frac{t_b^\star - 20}{20} \times 100\%, t_b^\star = \min \left\{ t \mid HV_{baselines}^{(t)} \ge HV_{\mathrm{SEETO}}^{(20)} \right\}$} represents the additional percentage of total evaluation steps required for the baselines to achieve the same HV value that SEETO attained at the $20$-th step. 
\revv{A task-wise Wilcoxon rank-sum test is conducted between SEETO and each baseline method. In the table, $\P$, $\dagger$, $\ddagger$, and $\S$ indicate that SEETO achieves a statistically significant advantage over MOTPE, MOASMO, KMO, and SAS-CKT-MO, respectively.}

\revv{In the high-similarity scenarios (Tasks~1--7), SEETO generally achieves higher HV values, demonstrating a clear early-stage search advantage. MOTPE underperforms SEETO in terms of both parameter search efficiency and overall performance. Compared with MOASMO and KMO, the superiority of SEETO indicates that historical task knowledge can effectively mitigate the "cold start" problem. In complex parameter calibration problems, MOASMO and KMO, which rely solely on the initial sampling and local surrogate models of the current task, find it difficult to rapidly identify high-quality parameter regions. SAS-CKT-MO narrows the gap with SEETO on this task, indicating that explicit knowledge transfer can improve search efficiency. Nevertheless, its performance remains inferior to that of SEETO, indicating that SEETO's knowledge transfer mechanism can exploit historical task information more effectively. 
For low-similarity tasks (Tasks~8--10), the advantage of SEETO over the baselines generally decreases. When the source task archive lacks historical tasks highly related to the target task, the effectiveness of transferable knowledge decreases, and the transfer gain becomes limited or may even introduce a search bias. Overall, SEETO significantly outperforms the baselines on most high-similarity tasks. On low-similarity tasks, its advantage decreases due to the reduced effectiveness of transferable historical knowledge, leading to comparable or only slightly better performance.}

\revv{Table~\ref{tab:hv20comparison} shows that, even under the 32 CPU cores parallel setting, WRF parameter calibration remains computationally expensive, requiring about 1.5 hours for 20 evaluations and nearly 5 hours for 60 evaluations. This supports the claim that WRF-based calibration is a computationally expensive optimization problem even when parallel computing resources are available. In addition, using 20 evaluations as the main reporting point is practically meaningful, since it roughly matches the acceptable time window in operational forecasting workflows, whereas 60 evaluations are better viewed as a supplementary budget for examining longer-term optimization behavior rather than a routine decision point in practice.}

Because MOTPE serves only as a non-evolutionary reference baseline and performs less favorably overall than the evolutionary methods, Fig.~\ref{fig:all_task_hv_line_colored} presents only the HV evolution trajectories of SEETO, KMO, MOASMO, and SAS-CKT-MO over 60 WRF evaluations. In the initial stage (1-20 evaluations), SEETO shows a clear warm-start advantage on high-similarity tasks (Tasks 1-7), because it can reuse historical knowledge from source tasks, whereas KMO and MOASMO have to build their surrogates from current-task data. On low-similarity tasks (Tasks 8-10), this advantage becomes smaller due to the lack of highly relevant source tasks. As the evaluations proceed to the intermediate stage (21-40 evaluations), KMO and MOASMO gradually accumulate target-task samples and improve their surrogate fitting to the current parameter landscape, while SAS-CKT-MO also continues to improve through competitive knowledge transfer and surrogate-assisted search. In contrast, the growth rate of SEETO slows down as it enters a fine-grained search stage. In the later stage (41-60 evaluations), SEETO maintains its lead on Tasks 1-7, confirming the robustness of knowledge transfer in related tasks. However, KMO and MOASMO gradually surpass SEETO and SAS-CKT-MO on Tasks 8-10, indicating that low-similarity source knowledge may cause negative transfer and guide the search toward suboptimal regions, thereby weakening final convergence performance.

\begin{figure*}[!t]
\centering
\includegraphics[width=7.0in]{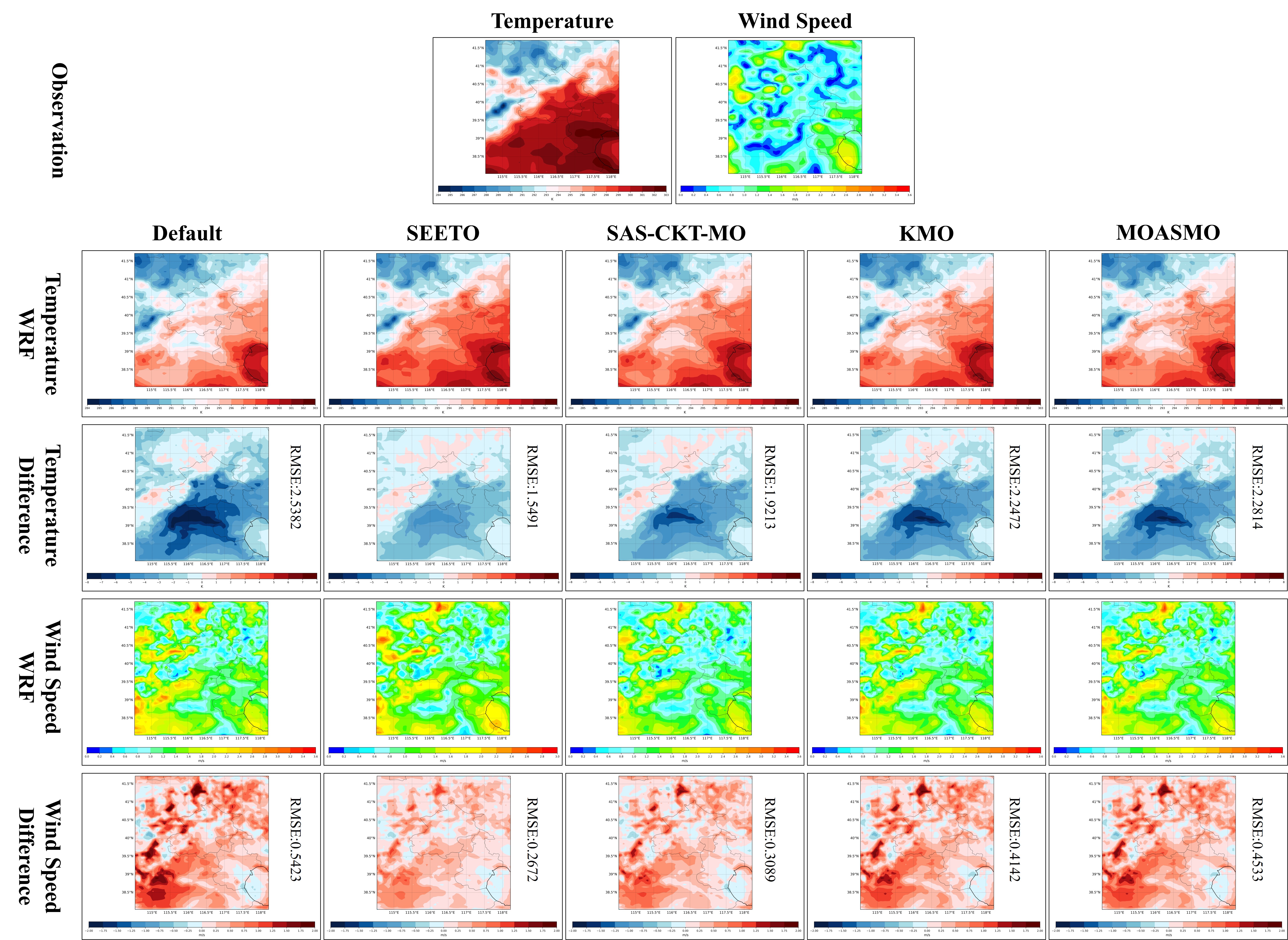}
\caption{\revv{Visual comparison of forecast fields and error maps for Task 1.}}
\label{fig:MetElementsplot}
\end{figure*}

\revv{Due to space limitations, Fig.~\ref{fig:MetElementsplot} provides a visual comparison for Task~1 after the complete 60 WRF evaluations, including the wind speed and temperature forecast fields generated by the optimal parameter vectors found by the commonly used WRF parameter calibration algorithms KMO, MOASMO and SAS-CKT-MO as well as SEETO, together with those generated by the WRF default parameter configuration. It also displays the corresponding wind speed and temperature difference fields between these forecasts and the true observations.}
First, the comparison of forecast fields in the first row reveals that while the default parameter configuration captures general meteorological trends, it exhibits visible deviations from ground truth in terms of local textures and extremum regions. Post-optimization, the forecast fields generated by SEETO, KMO, and MOASMO are all spatially closer to the true observations.
\revv{However, the performance disparities among the algorithms are more clearly discerned through the forecast difference maps in the second row and the corresponding RMSE values.} The difference map for the default parameters presents extensive dark regions (representing large positive/negative deviations), indicating significant systematic errors. 
Although KMO and MOASMO mitigate this bias to a certain extent, as evidenced by the lighter colors in the difference maps, distinct dark error patches persist in specific local regions. This implies that their calibrated parameters failed to completely eliminate local deviations. 
In contrast, the difference map corresponding to SEETO exhibits the most uniform and faintest hues, indicating the minimal residual magnitude between the forecast and observations. 
This implies that SEETO not only effectively rectifies the systematic bias of the default parameters but also adjusts physical parameters more precisely than KMO, MOASMO and SAS-CKT-MO to fit local micro-meteorological features, thereby achieving the highest forecast accuracy across the entire spatial domain.

\subsection{Ablation on Adaptive Knowledge Transfer Mechanism}
\revv{Our ablation study consists of two parts. First, we incorporate SEETO's knowledge transfer mechanism into KMO and MOASMO, resulting in two derivative variants, namely SEETO-KMO and SEETO-MOASMO, which are then compared with their respective original versions. Second, we simplify SEETO's own knowledge transfer mechanism to construct three transfer-based baseline variants for comparison, namely NTWS-SEETO, EIT-SEETO, and SET-SEETO.} \rev{The three transfer baselines keep the same evolutionary search process, surrogate-assisted selection mechanism, population size, evaluation budget, and relevant experimental parameter settings as SEETO, and differ only in whether the knowledge-transfer components are disabled or simplified. EIT-SEETO retains only solution-level transfer. It injects $n_{opt}=|\rho \cdot n_p|=20$ elite solutions selected from the retrieved source-task archives and fills the remaining 80 individuals randomly, but does not use the model-level ensemble surrogate transfer. EIT-SEETO is used to examine the effect of knowledge transfer when the model-level ensemble surrogate construction strategy is not adopted. SET-SEETO retains only model-level transfer. It constructs the multi-source ensemble surrogate but initializes the population without historical elite-solution injection. SET-SEETO is used to examine the effect of knowledge transfer when the solution-level elite injection strategy is not adopted. NTWS-SEETO retains both transfer levels but replaces SEETO's top-$\gamma$ source-task subset, with $\gamma=5$, by the single most similar source task, i.e., a top-1 transfer setting. NTWS-SEETO is used to examine the effect of using multiple similar source tasks as the knowledge-transfer sources compared with using only the single most similar source task.}

Fig.~\ref{fig:all_task_hv_line_colored} illustrates the HV evolution trajectories of the algorithms across all 10 tasks, comparing performance before and after the integration of the adaptive knowledge transfer mechanism. In high-similarity scenarios (Tasks 1-7), SEETO-KMO and SEETO-MOASMO significantly outperform their original counterparts, KMO and MOASMO. Particularly during the initial optimization phase, by reusing historical elite solutions and leveraging ensemble surrogate evaluation, these variants substantially improve the convergence starting point and search efficiency of the algorithms.
However, in low-similarity scenarios (Tasks 8-10), the inductive bias introduced by the adaptive knowledge transfer mechanism exerts a suppressive effect on algorithmic performance. The final convergence accuracy of SEETO-KMO and SEETO-MOASMO becomes inferior to that of the original KMO and MOASMO. This phenomenon indicates that when the transferred prior knowledge does not match the current optimization landscape, the forced injection of knowledge may bias the search direction, thereby triggering negative transfer.

\begin{figure*}[!t] 
    \centering
    \includegraphics[width=\textwidth]{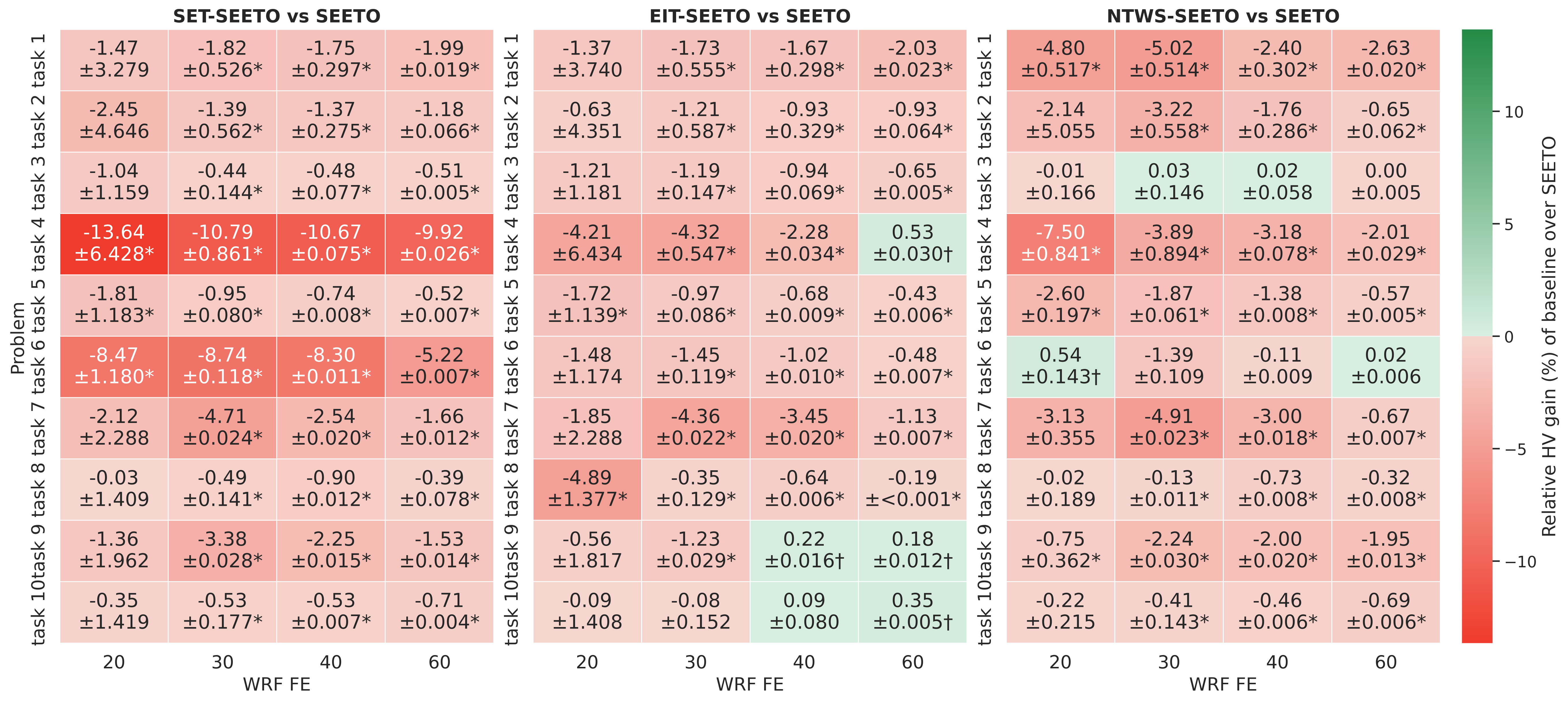}
    \caption{\revv{Relative HV gains of transfer baselines over SEETO.}}
    \label{fig:seeto_ablation}
\end{figure*}

\rev{Fig.~\ref{fig:seeto_ablation} compares the complete SEETO framework with three transfer-based baseline variants, namely SET-SEETO, EIT-SEETO, and NTWS-SEETO, across 10 sequential WRF calibration tasks and four function-evaluation budgets, i.e., FE = 20, 30, 40, and 60. For each task-budget-variant combination, we report the relative HV gain with respect to SEETO, defined as $\Delta HV(\%)=\frac{HV_{\text{baseline}}-HV_{\text{SEETO}}}{HV_{\text{SEETO}}}\times100\%.$ Each cell reports the mean ± standard deviation over 10 independent runs with different random seeds. Positive values indicate that the corresponding baseline variant achieves a higher HV than SEETO, whereas negative values indicate a higher HV for SEETO. To further assess the reliability of these differences, we performed paired Wilcoxon statistical tests between SEETO and each baseline variant under the same task and FE budget. In Fig. 6, * denotes cases where SEETO significantly outperforms the corresponding baseline, † denotes cases where the baseline significantly outperforms SEETO, and unmarked cells indicate that no significant difference was detected. Overall, negative values dominate Fig. 6. Across the 120 comparisons, SEETO achieves higher HV in 111 cases, including all 40 comparisons against SET-SEETO, 35 of 40 comparisons against EIT-SEETO, and 36 of 40 comparisons against NTWS-SEETO. The significance annotations further support this overall trend, while also showing that the magnitude and statistical reliability of the advantage vary across tasks, FE budgets, and ablated components. The few positive \(\Delta HV\) values mainly appear under larger FE budgets, where the initial advantage introduced by transfer becomes less pronounced as target-task feedback accumulates. These results are consistent with the design goal of SEETO: improving calibration efficiency under limited and expensive WRF evaluations, especially in the early optimization stage. Overall, the ablation results indicate that solution-level elite injection, model-level surrogate knowledge reuse, and multi-source task selection are complementary, and that SEETO’s advantage arises from their joint effect rather than from any single transfer component.}


\subsection{Comparative Analysis of Meteorological State Representation Extraction Methods}
To validate the effectiveness of the proposed meteorological state representation extractor in capturing task similarity and guiding transfer optimization, we conducted comparative experiments against two advanced self-supervised representation learning methods in the meteorological domain: SCL and AtmoDist. The final HV obtained in the WRF parameter calibration tasks serves as the primary evaluation metric.

\begin{table}[!t]
  \centering
  \caption{\revv{HV Comparison of Different Representation Learning Methods}}
  \label{tab:scl_atmodist_comparison}
  \footnotesize
  \setlength{\tabcolsep}{2pt}
  \renewcommand{\arraystretch}{1.08}
  \begin{tabular*}{\columnwidth}{@{\extracolsep{\fill}}lccc@{}}
    \toprule
    \multirow{2}{*}{Problem} & \multicolumn{3}{c}{HV} \\
    \cmidrule(lr){2-4}
     & \textbf{Ours} & SCL & AtmoDist \\
    \midrule
    \cellcolor{mygreen}Task 1  & \textbf{2.2214E-01}{\scriptsize$\pm$3.35E-04} & 2.2196E-01{\scriptsize$\pm$5.23E-04} & 2.1612E-01{\scriptsize$\pm$5.72E-04} \\
    \cellcolor{mygreen}Task 2  & \textbf{1.5314E-01}{\scriptsize$\pm$6.23E-04} & 1.5186E-01{\scriptsize$\pm$5.34E-04} & 1.5095E-01{\scriptsize$\pm$1.12E-03} \\
    \cellcolor{mygreen}Task 3  & \textbf{6.2880E-01}{\scriptsize$\pm$2.07E-04} & 6.2713E-01{\scriptsize$\pm$4.49E-04} & 6.2682E-01{\scriptsize$\pm$6.96E-04} \\
    \cellcolor{mygreen}Task 4  & \textbf{1.0639E-01}{\scriptsize$\pm$2.19E-04} & 1.0584E-01{\scriptsize$\pm$3.57E-04} & 1.0475E-01{\scriptsize$\pm$1.39E-04} \\
    \cellcolor{mygreen}Task 5  & \textbf{5.2233E-01}{\scriptsize$\pm$1.94E-04} & 5.2213E-01{\scriptsize$\pm$3.82E-04} & 5.2083E-01{\scriptsize$\pm$4.98E-04} \\
    \cellcolor{mygreen}Task 6  & \textbf{6.2817E-01}{\scriptsize$\pm$2.88E-04} & 6.2732E-01{\scriptsize$\pm$8.88E-04} & 6.2591E-01{\scriptsize$\pm$1.67E-04} \\
    \cellcolor{mygreen}Task 7  & \textbf{3.8624E-01}{\scriptsize$\pm$2.26E-04} & 3.8552E-01{\scriptsize$\pm$5.28E-04} & 3.8456E-01{\scriptsize$\pm$5.62E-04} \\
    \midrule
    \cellcolor{mygray}Task 8  & \textbf{5.1994E-01}{\scriptsize$\pm$2.64E-04} & 5.1524E-01{\scriptsize$\pm$7.20E-04} & 5.0133E-01{\scriptsize$\pm$6.64E-04} \\
    \cellcolor{mygray}Task 9  & \textbf{2.8571E-01}{\scriptsize$\pm$1.85E-04} & 2.8442E-01{\scriptsize$\pm$9.76E-04} & 2.8440E-01{\scriptsize$\pm$1.04E-04} \\
    \cellcolor{mygray}Task 10 & \textbf{5.3309E-01}{\scriptsize$\pm$1.33E-04} & 5.3255E-01{\scriptsize$\pm$1.51E-04} & 5.2982E-01{\scriptsize$\pm$1.63E-04} \\
    \bottomrule
  \end{tabular*}
\end{table}

Table~\ref{tab:scl_atmodist_comparison} presents the mean and standard deviation of the final HV values achieved by the three methods across 10 distinct target tasks. Experimental results indicate that the proposed method achieved superior performance across all tested tasks (Tasks 1–10). 
Compared to SCL, which is based on spatiotemporal contrastive learning, our method demonstrated a more pronounced advantage in complex tasks. 
This may be attributed to the fact that SCL primarily focuses on distinguishing categorical features of different weather systems. However, in parameter calibration tasks, capturing subtle spatial structural discrepancies within meteorological fields is of greater criticality. 
The performance of AtmoDist was relatively weaker. This is likely due to its pre-training objective, which focuses on predicting temporal intervals between atmospheric states. While effective at capturing dynamic evolutionary features, its ability to extract spatial features strongly correlated with physical parameters from static fields at a single time step appears somewhat limited. 
In contrast, the proposed method reconstructs meteorological states via an encoder-decoder architecture. This forces the latent representations to preserve more comprehensive spatial textures and physical consistency information, thereby providing more precise guidance for the subsequent transfer optimization process.


\subsection{Parameter Sensitivity Analysis}
To rigorously evaluate the robustness of SEETO with respect to key hyperparameters and to provide practical guidelines for parameter configuration, this subsection presents a sensitivity analysis focusing on four critical parameters: the number of channels representing the meteorological element dimensions, denoted as $\lambda$. The regulation parameter $c$, which governs the rate of transition from source domain knowledge to target domain local knowledge. \revv{The task similarity threshold coefficient $\tau$ for handling negative transfer. The temperature coefficient $\Gamma$, which controls the concentration degree of the Softmax weight distribution.}

The richness of the meteorological state representation directly dictates the accuracy of task similarity metrics. 
We categorized the input channel configurations for the feature extractor into three distinct levels for comparative analysis: 
a low-dimensional configuration (4 channels, comprising exclusively surface-level variables), a medium-dimensional configuration (19 channels, encompassing surface variables and upper-air variables across 3 key pressure levels), 
and a high-dimensional configuration (69 channels, including surface variables and the full spectrum of upper-air variables across 13 pressure levels). 
Fig.~S2 in the supplementary material illustrates the final HV performance (top panel) and the percentage of relative difference (bottom panel) under these varying channel configurations.
Observations from the relative difference plot reveal that, compared to the baseline (19 channels), the 4-channel configuration exhibits a performance degradation across the majority of tasks. 
Notably, in Task 4, the performance attenuation reaches as high as $7.88\%$. This suggests that relying solely on surface variables is insufficient to capture complex atmospheric dynamic features, consequently leading to biases in similar task retrieval. 
In contrast, although the 69-channel configuration yielded positive gains in most tasks, the magnitude of improvement in the remaining tasks was marginal. 
Considering that incorporating the full set of upper-air variables significantly increases data I/O load and the computational overhead of the feature extraction network, we conclude that the 19-channel configuration strikes the optimal balance between computational cost and representation effectiveness. 

The parameter $c$ controls the transition rate of surrogate-model dominance from the source-task ensemble model to the target-domain local model. A larger $c$ slows this transition and allows the algorithm to exploit source-domain prior knowledge for longer, whereas a smaller $c$ makes the search rely earlier on the surrogate model built from target-task data. Fig.~S3 in the supplementary material shows the HV trends of SEETO across 10 target tasks when $c \in \{0.017,0.024,0.031,0.038\}$. The results indicate that the preferred value of $c$ is closely related to source-target task similarity: for high-similarity tasks (Tasks 1-7), a larger $c$ better preserves useful source knowledge and improves search performance. For low-similarity tasks (Tasks 8-10), a smaller $c$ more rapidly weakens irrelevant source-domain influence and shifts the search toward target-task feedback, thereby mitigating negative transfer.

\rev{Fig.~S4 in the supplementary material shows the HV variation of SEETO across 10 target tasks when $\Gamma$ is set to $0.025$, $0.05$, $0.065$, $0.10$, and $0.20$. In the current WRF calibration scenario, the default setting $\Gamma=0.065$ achieves the best average HV among all tested values. A smaller $\Gamma$ makes the Softmax weights overly concentrated on a few source tasks with the highest similarities, thereby weakening the complementary information among multiple source tasks, whereas a larger $\Gamma$ makes the weight distribution overly smooth and weakens the guidance from highly similar source tasks. Under the current 10 target-task scenarios, $\Gamma=0.065$ achieves a better balance between emphasizing highly similar source tasks and preserving multi-source complementary information.}

\revv{Fig.~S5 in the supplementary material presents the sensitivity results when $\tau$ is set to $0.5$, $0.6$, $0.65$, $0.7$, and $0.8$. This parameter is used to balance historical knowledge exploitation and negative transfer suppression. The results show that the default setting $\tau=0.7$ achieves the best overall average HV across the 10 tasks, indicating a favorable trade-off between transfer gains and negative transfer control. Lower thresholds, i.e., $\tau=0.5$, $0.6$, and $0.65$, generally lead to weaker performance, suggesting that overly relaxed transfer admission criteria introduce unreliable source task knowledge. In contrast, although the higher threshold $\tau=0.8$ can alleviate negative transfer in some low-similarity tasks, its overall performance is still inferior to that of $\tau=0.7$, because it weakens the exploitation of effective historical knowledge in high-similarity tasks and reduces the early-stage search advantage brought by transfer.}

\section{Conclusion}
\label{Section-V}
To address the challenges of prohibitive computational costs and the inefficiency associated with repetitive optimization from scratch in multi-task NWP model parameter calibration, this paper proposes SEETO.
This framework transcends the traditional assumption that calibration tasks are mutually independent. By constructing a knowledge repository of historical tasks, \revv{it enables the use of historical task knowledge to reduce repeated optimization from scratch.}

\revv{In high-similarity scenarios, SEETO can exploit historical knowledge more effectively, exhibiting a clear early-stage convergence advantage and achieving favorable performance in both the HV metric and spatial forecast accuracy. In contrast, in low-similarity scenarios, although SEETO usually retains a certain initial advantage, negative transfer caused by the mismatch between source task knowledge and the target-task landscape may emerge in the later stage of optimization. This suggests that relying solely on static similarity metrics is still insufficient to fully accommodate the dynamically changing search requirements during optimization, and more adaptive strategies for dynamically avoiding negative transfer are needed. Meanwhile, the current method still has two limitations. First, due to the high computational cost of WRF evaluations, the number of independent runs and calibration tasks is inevitably limited; therefore, the experimental results should be interpreted within the current experimental scale. Second, this work mainly focuses on WRF parameter calibration under a specific configuration, and whether SEETO can be generalized to other physical scheme configurations, more calibration objectives, and broader expensive sequential calibration scenarios remains to be further investigated in future work.}

\bibliographystyle{IEEEtran}
\nocite{*}  
\bibliography{references}

\vspace{11pt}

\vfill

\clearpage
\includepdf[
  pages=-,
  pagecommand={}
]{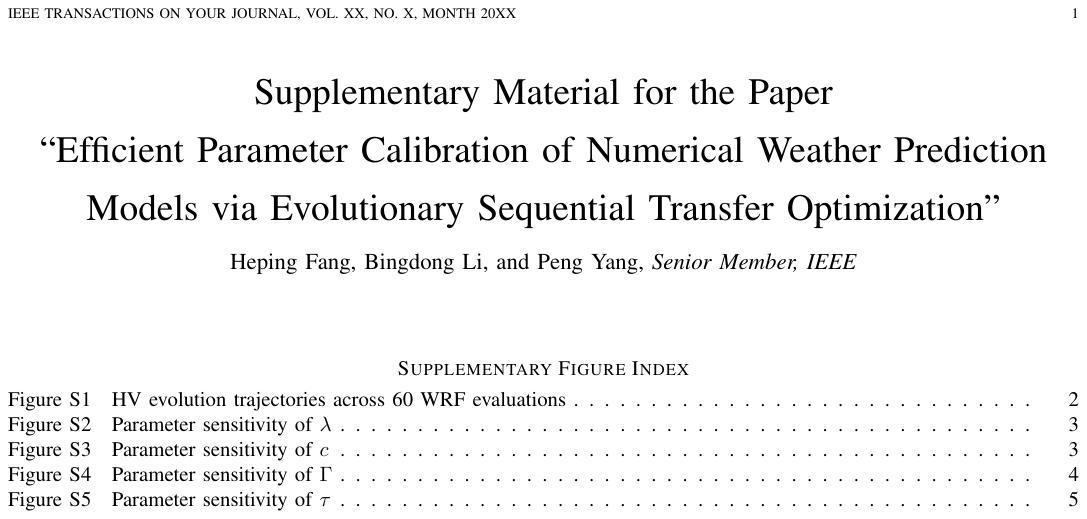}

\end{document}